\documentclass[aps,pra,groupedaddress,superscriptaddress,bibnotes,amsfonts,longbibliography,citeautoscript,a4paper,10pt,nohyperref]{revtex4-2}

\usepackage{latexsym}
\usepackage{amsmath}
\usepackage{amssymb}
\usepackage{graphicx}
\usepackage[T1]{fontenc}
\usepackage[open]{bookmark}
\usepackage{hyperref}
\usepackage{orcidlink}
\hypersetup{colorlinks=true,allcolors=blue}
\usepackage{xcolor} 
\usepackage{comment}
\usepackage{bm}
\newcommand{\vecnabla}{\bm{\nabla}}
\providecommand{\vect}[1]{{\mathbf{#1}}}
\DeclareRobustCommand{\uvec}[1]{{
  \ifcsname uvec#1\endcsname
     \csname uvec#1\endcsname
   \else
    \bm{\hat{\mathbf{#1}}}%
   \fi
}}

\newcommand{\elc}{e}
\newcommand{\ii}{\mathrm{i}}
\newcommand{\ex}[1]{\mathrm{e}^{#1}}
\newcommand{\utheta}{\bm{\hat{\uptheta}}}

\usepackage{upgreek}
\newcommand{\uphi}{\bm{\hat{\upphi}}}
\newcommand{\urho}{\bm{\hat{\uprho}}}
\renewcommand{\eqref}[1]{Eq.~(\ref{#1})}
\newcommand{\eqsref}[1]{Eqs.~(\ref{#1})}

\newcommand{\dd}{\textrm{d}}
\newcommand{\eye}{\mathbb{I}}
\newcommand{\zero}{\mathbb{O}}

\usepackage{xspace}
\newcommand{\treams}{\textit{treams}\xspace}
\newcommand{\treamsebeam}{\textit{treams\_ebeam}\xspace}

\usepackage{listings}
\begin{document}

\title{T-matrix scattering formalism for electron-beam spectroscopy}

\author{P.~Elli~Stamatopoulou\,\orcidlink{0000-0001-9121-911X}}
\email{elli.stamatopoulou@kit.edu}
\affiliation{Institute of Nanotechnology, Karlsruhe Institute of Technology, Kaiserstr. 12, DE-76131, Germany}
\affiliation{POLIMA---Center for Polariton-driven Light–-Matter Interactions, University of Southern Denmark, Campusvej 55, DK-5230 Odense M, Denmark}
\author{Ulrich~Hohenester\,\orcidlink{0000-0001-8929-2086}}
\affiliation{Institute of Physics, University of Graz,
Universit\"atsplatz 5, AT-8010, Austria}
\author{Carsten~Rockstuhl\,\orcidlink{0000-0002-5868-0526}}
\email{carsten.rockstuhl@kit.edu}
\affiliation{Institute of Nanotechnology, Karlsruhe Institute of Technology, Kaiserstr. 12, DE-76131, Germany}
\affiliation{Institute of Theoretical Solid State Physics, Karlsruhe Institute of Technology, Kaiserstr. 12, DE-76131, Germany}
\affiliation{Center for Integrated Quantum Science and Technology, Karlsruhe Institute of Technology, Wolfgang-Gaede-Str. 1, 76131 Karlsruhe, Germany}

\date{\today}

\begin{abstract}
Advanced computational tools that describe the interaction of electrons with structured nanophotonic devices are crucial for theoretical predictions, specific design tasks, and the interpretation of experimental results. These tools open the door to a systematic exploration of light--matter interactions in complex photonic environments and support the development of next-generation nanophotonic platforms. Here, we report on the implementation of electron-beam spectroscopy in a T-matrix-based scattering formulation. Such a framework is quite versatile in predicting the electromagnetic response of complex photonic materials composed of periodically or aperiodically arranged individual scatterers. By extending this formalism to describe interactions with fast electrons, we provide a fast and accurate numerical tool for simulating cathodoluminescence (CL) and electron energy-loss spectroscopy (EELS) measurements. The desired functionalities are implemented in \treamsebeam, an extension of the existing software suite \treams for electromagnetic scattering computations, available at \url{https://github.com/tfp-photonics/treams_ebeam}. We demonstrate the implementation details on a carefully selected set of problems, including single scatterers of various shapes and materials, a periodic chain of elliptical nanodisks, and a finite cluster of nanospheres arranged in a two-dimensional (2D) lattice. By uniting fast-electron physics with advanced scattering theory, our framework unlocks new possibilities for designing, understanding, and engineering next-generation nanoscale light--matter interactions.
\end{abstract}

\maketitle

\section{Introduction}\label{sec:introduction}

In recent years, electron-beam spectroscopy has proven indispensable for exploring the optical properties of matter. Fast electrons passing by a sample transfer energy to the optical modes sustained therein, which can then be detected using two complementary techniques~\citep{GarciadeAbajo:2010rmp}. In cathodoluminescence (CL) spectroscopy, a tightly focused electron beam is scanned over a specimen in an electron microscope, while the far-field emission released from the sample is collected and analyzed to reveal the radiative decay pathways of its optical excitations~\cite{kociak_um176,coenen_apr4}. Electron energy-loss spectroscopy (EELS), performed in the transmission configuration of the electron microscope, instead detects both radiative and dark optical modes~\cite{egerton:2011}. Remarkable progress in theoretical and experimental approaches in both techniques has led to an unprecedented combination of spatial, energy, and—most recently—temporal resolution~\cite{Polman:2019,GarciadeAbajo:2021acsp}.

Unlike in traditional light-based spectroscopic methods, fast electrons can access the near field of a photonic structure because the electromagnetic fields accompanying them contain evanescent spectral components. At the same time, they are spectrally broadband, and can excite different collective phenomena across a wide range of frequencies, from bulk and surface plasmons in metals~\cite{watanabe:1956,ritchie:1957,Yamamoto:2001,Vesseur:2007} and Mie resonances in dielectric nanocavities~\cite{coenen:2013,matsukata:2019,Fiedler:2022} in the optical spectrum, to THz bulk and surface phonons in polar crystals~\cite{lagos:2017,hage:2019,govyadinov_nc8,maciel_prb102} and magnons in magnetic materials~\cite{kepaptsoglou_nat644}. Owing to these characteristics, fast electrons are powerful probes for studying more complex photonic platforms designed for greater optical control, such as nanoantennas, photonic crystals, or metasurfaces, which comprise multiple scatterers, often in periodic arrangements.
Such interactions have already been investigated, though not extensively, for the generation of Smith--Purcell~\cite{smith_purcell, pendry_martinmoreno,chen_ol48} and Cherenkov radiation~\cite{garciadeabajo_prl91}, or bound (BICs) and quasi-bound states in the continuum (q-BICs)~\cite{dong_lsa11}.
    
As photonic platforms grow in complexity, so does the need for robust numerical tools capable of modeling the free electron--matter interaction efficiently. In this regard, several methods have been devised for CL and EELS simulations. These include the boundary element method (BEM)~\cite{GarciadeAbajo:2002prb,hohenester:2014}, the finite element method (FEM)~\cite{maciel_nc14,dong_lsa11}, the finite-difference time-domain (FDTD) method~\cite{das:2012,cao:2015}, the discrete dipole approximation (DDA)~\cite{kichigin_jpcc127}, or the discontinuous Galerkin time-domain (DGTD) method~\cite{matyssek:2011,Husnik:2013,Pramassing:2021, stamatopoulou_prr6}. While these approaches undeniably have their merits, they rely on solving Maxwell's equations in discretized space and/or time, rendering them computationally expensive and time-consuming, even prohibitive for systems involving multiple length scales or requiring extensive parameter sweeps. Furthermore, modeling infinitely extended periodic structures remains a challenge for some of these methods. 

An alternative approach that overcomes these limitations is the T-matrix method~\cite{waterman_piee53,mishchenko_jqsrt111,mishchenko_jqsrt88}. In this framework, the fields scattered by an object are related to the illumination via the T-matrix, which encodes all material and geometrical parameters of the entire structure, and is otherwise independent of the illumination. The computational benefit comes from the fact that the T-matrix only needs to be retrieved once for a given object, and can be stored and reused for later computations~\cite{asadova_jqsrt333,asadova:2026}. Moreover, the response to electron beams from complex photonic materials made from many scatterers can conveniently be expressed within an algebraic formulation~\cite{mackowski_josaa13,xu_ao34,fernandez_corbaton_apr4,rasoul_aom7}. Early theoretical efforts in electron-beam spectroscopy employed the T-matrix method to study the individual and collective resonances in spherically and cylindrically symmetric nanoparticles and assemblies thereof ~\cite{GarciadeAbajo:1999prb,garciadeabajo_pre61,zabala:1989,walsh_pmb63,rodriguez_prb111}.

Here, we present a general framework of CL and EELS simulations based on the T-matrix method, applicable to one or more scatterers of arbitrary shapes and electron beams whose trajectories do not intersect the structures. The manuscript is structured as follows. We first outline the theory, in which we express the field of a fast electron in the appropriate basis so that it can be considered in the T-matrix formalism. We then focus on its interaction with a scatterer, demonstrating the relation between the scattered and incident field via the T-matrix formalism, which we extend to account for multiple scatterers arranged periodically or aperiodically. Based on this formalism, we show the resulting expressions for computing the CL and EELS probabilities. In the following section, we demonstrate the capabilities of our approach through three representative examples: single scatterers of various shapes, a periodic chain of elliptical nanodisks, and a finite cluster of nanospheres arranged in a two-dimensional (2D) lattice. We present our numerical implementation of this framework in \treamsebeam, an extension of the Python package \treams~\cite{beutel_cpc297}, providing guidance on its use. To demonstrate the computational advantages and limitations of the proposed approach, we analyze its computational requirements and compare its performance against an alternative method.


\section{Theory}\label{sec:theory}

This section outlines the theoretical framework underlying our description of the interaction of electrons with structured nanophotonic materials, progressing from the field representation of a fast electron to experimentally accessible observables. We first formulate the electromagnetic field of a relativistic electron in a cylindrical wave basis (CWB), which provides a natural and efficient representation for an electron moving along a straight trajectory. We then describe the interaction of the electron with complex photonic nanostructures composed of isolated scatterers and of ensembles thereof, arranged periodically or aperiodically. Eventually, we show how the same formalism yields unified expressions for CL and EELS probabilities, directly linking theory to measurable radiation and dissipation channels.

\subsection{Electromagnetic field of a fast electron in the cylindrical wave basis}
\label{subsec:theory:ebeam}

We start by expressing the electromagnetic fields of an electron traveling with constant velocity in the angular-frequency ($\omega$) domain in the CWB, using a cylindrical coordinate system $\vect{r} =(\rho, \phi, z)$. The choice of basis is motivated by the fact that the fields of the electron take the form of a transverse magnetic (TM) cylindrical wave, as we demonstrate in what follows. We note here that all relevant fields in the present T-matrix formalism assume electron beams propagating in vacuum with relative permittivity and permeability equal to unity.

The electric field $\vect{E} ({\vect{r}, \omega})$ of an electromagnetic wave can be written in the following general form in the CWB~\cite{Bohren_Wiley1983}:
\begin{equation} \label{eq:efield_cwb}
    \vect{E} ({\vect{r}, \omega})= \sum_{m=-\infty}^{+\infty} \int_{-\infty}^\infty \dd k_z \left[  b_{k_z m} \vect{M}^{(n)}_{k_z m} ({\vect{r}, \omega}) + a_{k_z m} \vect{N}^{(n)}_{k_z m} ({\vect{r}, \omega})  \right] \, ,
\end{equation}
where $\vect{M}^{(n)}_{k_z m}({\vect{r}, \omega})$ and $\vect{N}^{(n)}_{k_z m}({\vect{r}, \omega})$ are vector cylindrical wave functions of kind $n$, describing vector cylindrical waves of transverse electric (TE) and transverse magnetic (TM) polarization, respectively. They are characterized by wavenumber $k(\omega)=\omega/c$, where $c = 1/\sqrt{\varepsilon_0\mu_0}$ is the speed of light in vacuum, and $\varepsilon_0$, $\mu_0$ are the vacuum permittivity and permeability, respectively. The vector cylindrical waves are indexed by the $z$-component of their wavevector $k_z$ and the azimuthal index $m \in \mathbb{Z}$. Furthermore, $ b_{k_z m}$ and $ a_{k_z m} $ are the corresponding expansion coefficients that express the contribution of a given vector cylindrical wave to the electric field. The basis functions of the vector cylindrical waves in a cylindrical coordinate system characterized by the basis vectors $\urho$, $\uphi$, and $\uvec{z}$ are given by the following relations:
\begin{subequations} \label{eq:cwb_vectors}
    \begin{align}
   \vect{M}^{(n)}_{k_z m} ({\vect{r}, \omega}) &=  \left(\frac{\ii m}{k_{\rho} \rho} Z_{m}^{(n)} (k_{\rho} \rho) \urho - {Z_{m}^{(n)}}' (k_{\rho} \rho) \uphi \right) \ex{\ii m \phi + \ii k_z z} \, ,\\
   \vect{N}^{(n)}_{k_z m} ({\vect{r}, \omega}) &=  \left(\frac{\ii k_z}{k} {Z_{m}^{(n)}}' (k_{\rho} \rho) \urho - \frac{k_z m}{k k_{\rho}\rho} Z_{m}^{(n)} (k_{\rho} \rho) \uphi + \frac{k_{\rho}}{k}  Z_{m}^{(n)} (k_{\rho} \rho) \uvec{z} \right) \ex{\ii m \phi + \ii k_z z} \, , 
\end{align}
\end{subequations}
where $k_{\rho} (\omega)= \sqrt{k^2(\omega)-k_z^2}$ is the wavevector component in the radial direction. For notational simplicity, in \eqsref{eq:cwb_vectors} and in what follows, the explicit dependence of  $k$ and $k_\rho$ on $\omega$ is omitted. The function $Z_m^{(n)}(x)$ can be the Bessel or Hankel function of the first or second kind, respectively, and ${Z_{m}^{(n)}}' (x)$ denotes its derivative with respect to $x$. As usual for T-matrices, we choose $Z_m^{(1)}(x) = J_m(x)$ to describe waves that must be regular in the whole space, and $Z_m^{(3)}(x) = H^{(+)}_m(x)$ for those that must fulfill the radiation condition. Henceforth, we refer to these two types of waves as \textit{regular} and \textit{singular}, respectively.

We now consider an electron carrying the negative elementary charge $-\elc$ and traveling along the $z$~axis with constant velocity $\vect{v} =v\uvec{z}$, such that its trajectory is given by $\vect{r}_e = \vect{v}t$. At time $t=0$, the electron lies at the origin of the axes. The electric field for this case can be found in the literature~\cite{stamatopoulou_josab42}, and it reads:
\begin{equation} \label{eq:ebeam_efield}
    \vect{E} ({\vect{r}, \omega})= -\frac{\elc}{2\pi\varepsilon_0}  \frac{\omega}{v^2\gamma} \ex{\ii \omega z/v} \left[ K_1 \left(\frac{\omega \rho}{v\gamma}\right) \urho - \frac{\ii}{\gamma} K_0 \left(\frac{\omega \rho}{v\gamma}\right) \uvec{z} \right]\, ,
\end{equation}
where $\gamma = 1/\sqrt{1-\beta^2}$ is the Lorentz boost, with $\beta = v/c$ being the beta factor. Here, $K_m(x)$ denotes the modified Bessel function of order $m$. To express the field in the basis of \eqref{eq:efield_cwb}, we apply the connection formulas $K_m (x) =\pi \ii^{m+1} H_m^{(+)} (\ii x)/2$~\cite{arfken_weber}, and rewrite \eqref{eq:ebeam_efield} as
\begin{equation}
     \vect{E} ({\vect{r}, \omega})= \frac{\elc}{4\varepsilon_0}  \frac{\omega}{v^2\gamma} \ex{\ii \omega z/v} \left[ H^{(+)}_1 \left(\frac{\ii\omega \rho}{v\gamma}\right) \urho - \frac{1}{\gamma} H^{(+)}_0 \left(\frac{\ii\omega \rho}{v\gamma}\right) \uvec{z} \right]\, .
\end{equation}
As a next step, we use the property of Hankel functions $H^{(+)}_1 (x) = -{H_0^{(+)}}' (x)$~\cite{arfken_weber}, and find
\begin{equation}
    \vect{E} ({\vect{r}, \omega})= -\frac{\elc}{4\varepsilon_0}  \frac{\omega}{v^2\gamma} \ex{\ii \omega z/v} \left[ {H_0^{(+)}}' \left(\frac{\ii\omega \rho}{v\gamma}\right) \urho + \frac{1}{\gamma} H^{(+)}_0 \left(\frac{\ii\omega \rho}{v\gamma}\right) \uvec{z} \right]\, .
\end{equation}
By introducing $k_z = \omega/v$ and $k_{\rho} = \ii \omega/ (v\gamma)$, with $\sqrt{k_\rho^2+k_z^2} = \omega/c = k$, we obtain
\begin{equation}\label{eq:ebeam_efield_cwb}
    \vect{E} ({\vect{r}, \omega})= \frac{\ii \elc}{4\varepsilon_0}  \frac{k_z}{c\gamma} \ex{\ii k_z z} \left[ \frac{\ii k_z}{k}  {H_0^{(+)}}' \left(k_{\rho}\rho\right) \urho + \frac{k_{\rho}}{k} H^{(+)}_0 \left(k_{\rho}\rho\right) \uvec{z} \right]\, .
\end{equation}
Compared with \eqsref{eq:efield_cwb} and (\ref{eq:cwb_vectors}), \eqref{eq:ebeam_efield_cwb} reveals that the field produced by the electron is a purely TM-polarized cylindrical wave of singular type, order $m=0$ and fixed $k_z$, whose amplitude is given by $a_{k_z 0} = \ii \elc k_z/(4\varepsilon_0c\gamma)$. Furthermore, if we express $k_z$ in terms of the wavenumber, we find $k_z = k/\beta>k$ and $k_{\rho}= \ii k/(\beta\gamma) \in \mathbb{I}$, with $\mathbb{I}$ denoting the set of imaginary numbers, indicating that the cylindrical wave decays exponentially away from its axis. Interestingly, the excitation of a nanostructure by a fast electron is one of the few electromagnetic scattering problems that is often best formulated in terms of cylindrical rather than spherical or plane waves. In the following section, we focus on this interaction and discuss in more detail the choice of basis for the expansion of the relevant fields.

\subsection{Interaction of a fast electron with nanostructures}
\label{subsec:theory:ebeam_nanostructure}

\begin{figure}[t]
    \centering
    \includegraphics[width=0.7\linewidth]{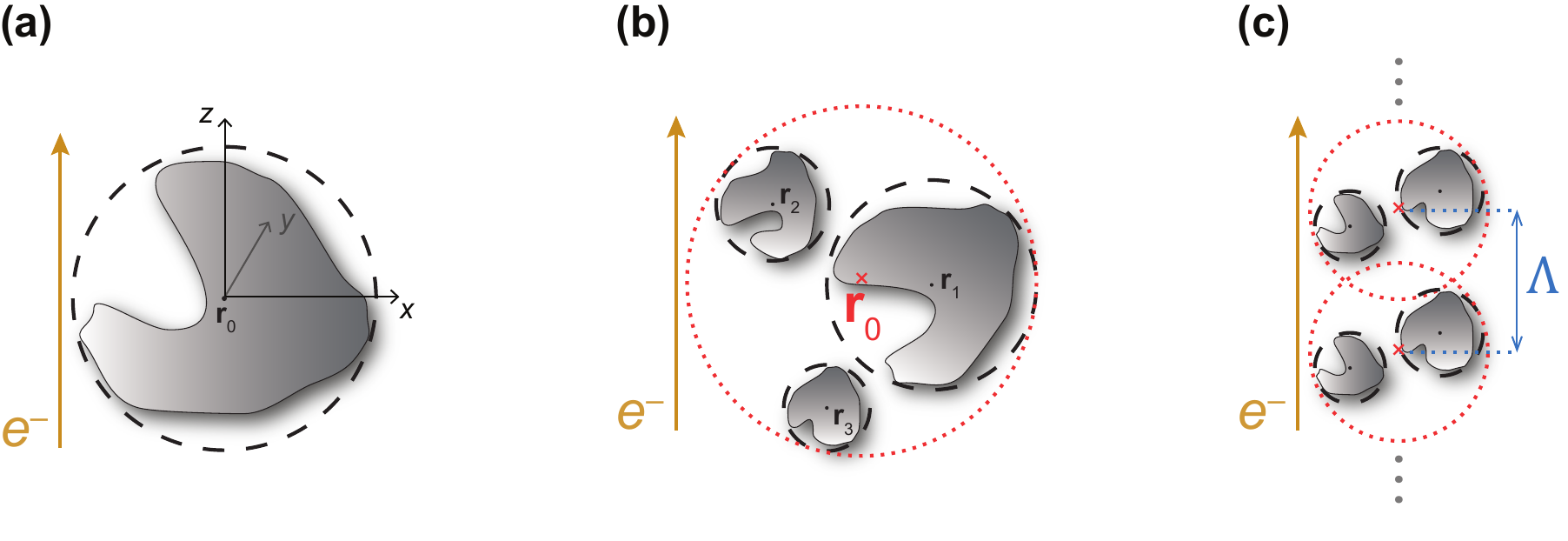}
    \caption{Examples of scatterer configurations excited by an electron beam, along with the validity domains of the relevant field expansions in the SWB. (a)~For a scatterer of arbitrary shape, the validity domain is outside the circumscribing sphere enclosing the scatterer, outlined with the dashed black circle.
    (b)~Cluster with three scatterers. In a local description, the fields are expanded at the position $\vect{r}_i$ of the $i$th scatterer, and are valid outside the corresponding circumscribing sphere (black dashed circle). The red dotted circle indicates the validity domain of the global description, which lies outside the circumscribing sphere enclosing all scatterers.   
    (c)~Periodic 1D array with pitch $\vect{\Lambda} = \Lambda \uvec{z}$ with a complex unit cell, composed of two scatterers. The red dotted circles mark the domain of validity of a global T-matrix, which describes the unit cell scatterers as a single object. This description might be invalid due to overlapping circumscribing spheres, in which case only the local description (black dashed circles) can be employed.}
    \label{fig:domains}
\end{figure}

This section summarizes the basic formulations of the T-matrix method for electromagnetic scattering computations, including multiple scattering scenarios and periodic boundary conditions. We first consider the interaction of a fast electron with a single scatterer of arbitrary shape, as shown in Fig.~\ref{fig:domains}(a). The incident field of the electron excites the optical modes of the object, leading to an additional scattered field. The incident and scattered fields, which constitute the total field, can be expanded about the position of the scatterer as
\begin{subequations}
\label{eq:ebeam_efield_eqs}
    \begin{align}
    \label{eq:ebeam_efield_general}
        \vect{E}_\mathrm{inc} ({\vect{r}, \omega}) &= \sum_{\nu} \sum_{s=0, 1} a_{\nu s} \vect{F}_{\nu s}^{(1)} \left({\vect{r}, \omega}\right)\, , \\
    \label{eq:scat_efield_general}
        \vect{E}_\mathrm{sca} ({\vect{r}, \omega}) &= \sum_{\nu} \sum_{s=0, 1} p_{\nu s} \vect{F}_{\nu s}^{(3)} \left({\vect{r}, \omega}\right)\, ,
    \end{align}
\end{subequations}
where $\nu$ is a generalized index condensing all other indices used to expand the field, $s$ is a polarization index, and $\vect{F}_{\nu s}^{(n)}\left({\vect{r}, \omega}\right)$ are the basis vector functions. To connect this notation with the formulas in Section~\ref{subsec:theory:ebeam}, in the CWB we have $\nu = k_z, m$, 
while $\vect{F}_{\nu 0}^{(n)}\left({\vect{r}, \omega}\right) = \vect{M}_{k_z m}^{(n)}({\vect{r}, \omega})$ for TE polarization ($s= 0$), and $\vect{F}^{(n)}_{\nu 1}\left({\vect{r}, \omega}\right) = \vect{N}^{(n)}_{k_z m}({\vect{r}, \omega})$ for TM polarization ($s= 1$). In the spherical wave basis (SWB), $\nu = \ell, m$, 
while $\vect{F}_{\nu 0}^{(n)}\left({\vect{r}, \omega}\right) = \vect{M}_{\ell m}^{(n)}({\vect{r}, \omega})$ for TE polarization, and $\vect{F}^{(n)}_{\nu 1}\left({\vect{r}, \omega}\right) = \vect{N}^{(n)}_{\ell m}({\vect{r}, \omega})$ for TM polarization, with $\ell, m$ denoting the angular momentum indices with $-\ell \leq m \leq \ell$. In a spherical coordinate system $(\rho, \theta,\phi)$ with unit vectors $\uvec{r}$, $\utheta$, $\uphi$, the basis vectors of the electric fields in the SWB take the general form~\cite{Bohren_Wiley1983}
\begin{subequations} \label{eq:swb_vectors}
    \begin{align}
   \vect{M}^{(n)}_{\ell m} ({\vect{r}, \omega}) &= N_{\ell m} \left[\ii \pi_{\ell m} (\theta) \utheta - \tau_{\ell m} (\theta) \uphi \right] \ex{\ii m\phi} =  \vect{X}_{\ell m} (\theta, \phi) z_{\ell}^{(n)} (kr) \, , \\
   \vect{N}^{(n)}_{\ell m} ({\vect{r}, \omega}) &=  \left[ \ii \sqrt{\ell (\ell +1)} Y_{\ell m} (\theta, \phi) \frac{z_{\ell}^{(n)} (kr)}{kr} \uvec{r} + \left(z_{\ell}^{(n)'}(kr) + \frac{z_{\ell}^{(n)} (kr)}{kr} \right)\uvec{r} \times \vect{X}_{\ell m } (\theta, \phi)  \right] \, ,
\end{align}
\end{subequations}
where
\begin{align}
    N_{\ell m}  = \ii\sqrt{\frac{(2\ell +1)}{4\pi \ell (\ell+1)}\frac{(\ell -m)!}{(\ell+m)!}} \, , \quad
    \pi_{\ell m} (\theta) = \frac{m P_{\ell}^m (\cos{\theta})}{\sin{\theta}} \, , \quad
    \tau_{\ell m} (\theta) = \frac{\partial P_{\ell}^m (\cos{\theta})}{\partial{\theta}} \, .
\end{align}
In \eqsref{eq:swb_vectors}, $Y_{\ell m}(\theta, \phi)$ are the scalar spherical harmonics, $\vect{X}_{\ell m}(\theta, \phi)$ are the vector spherical harmonics, and the angular functions $\pi_{\ell m}(\theta)$ and $\tau_{\ell m}(\theta)$ are derived from the Legendre polynomials $P_{\ell}^m (\cos{\theta})$. The radial dependence of the basis vectors is given by spherical Bessel-type functions, where $z_{\ell}^{(1)}(x) = j_{\ell}(x)$ are spherical Bessel functions of the first kind (regular modes) and $z_{\ell}^{(3)}(x) = h^{(+)}_{\ell}(x)$ spherical Hankel functions of the first kind (singular modes). For all formulas and basis vectors presented here, we use the conventions found in Ref.~\citenum{beutel_cpc297}. We emphasize that the spherical-wave expansion of the fields given in \eqref{eq:ebeam_efield_eqs} holds in homogeneous space. As a result, it is valid only outside the circumscribing sphere enclosing the scatterer, as indicated in Fig.~\ref{fig:domains}(a) with the dashed black circle. As a consequence of this limitation, an electron beam intersecting the invalid domain, including those that penetrate the scatterer, cannot be described by \eqref{eq:ebeam_efield_general}.Various approaches have been devised to circumvent this limitation~\cite{lamprianidis_jqsrt296, egel_oe24, bertrand_josaa37}, albeit at the cost of higher computational complexity and resources.

In Section~\ref{subsec:theory:ebeam}, we expressed the field of the electron in terms of singular waves [\eqref{eq:ebeam_efield_cwb}] around the electron position at time $t=0$, whereas in \eqref{eq:ebeam_efield_general} it is expanded in terms of regular waves around the position of the scatterer. To convert a singular cylindrical wave at point $\vect{r}$ to a regular one at $\vect{r}'$,we use the expression~\cite{huang_jmmct4}
\begin{equation} \label{eq:CW_to_CW}
    \vect{F}^{(3)}_{k_z m \,s} (\vect{r}, \omega) = \sum_{m'=-\infty}^{\infty} H_{m-m'}^{(+)} \left( \sqrt{k^2 - k_z^2} \rho_{r-r'}\right) \ex{\ii (m-m')\phi_{r-r'} + \ii k_z (z-z')} \,\vect{F}^{(1)}_{k_z m' s} (\vect{r}', \omega) \,.
\end{equation}
Additionally, we can expand the incident field of a fast electron in the SWB using the relations connecting regular cylindrical and regular spherical waves, given by~\cite{han_joapao10} 
\begin{align} \label{eq:CW_to_SW}
    \begin{pmatrix}
        \vect{M}_{k_z m}^{(1)} (\vect{r}, \omega)\\
        \vect{N}_{k_z m}^{(1)} (\vect{r}, \omega)
    \end{pmatrix}
    =
    \sum_{\ell = \mathrm{max} (|m|,1)}^{\infty} 4\pi \ii^{\ell-m} N_{\ell m}\ex{\ii m \phi_k}
    \begin{pmatrix}
        \tau_{\ell m} (\theta_k) & \pi_{\ell m} (\theta_k)\\
        \pi_{\ell m} (\theta_k) & \tau_{\ell m} (\theta_k)
    \end{pmatrix}
    \begin{pmatrix}
        \vect{M}_{\ell m}^{(1)} (\vect{r}, \omega)\\
        \vect{N}_{\ell m}^{(1)} (\vect{r}, \omega)
    \end{pmatrix} \, ,
\end{align}
where $\theta_k$ and $\phi_k$ are the angles of the wavevector $\vect{k} = (k, \theta_k, \phi_k)$ in spherical coordinates. 

Returning to \eqsref{eq:ebeam_efield_eqs}, the scattered and incident field in a linear response theory are related via a T-matrix as~\cite{mishchenko_2002}
\begin{equation} \label{eq:p_single}
    \vect{p} = \vect{T} \, \vect{a}\, ,
\end{equation}
where the vectors $\vect{p}$ and $\vect{a}$ are composed of the respective scattered and incident field coefficients. The T-matrix of an object encodes all material and geometrical properties that determine its optical response to an illumination. To perform the matrix multiplication of \eqref{eq:p_single}, both the T-matrix and the illumination must be represented in the same basis. Whether the object is best described by a spherical or a cylindrical T-matrix depends on its symmetry. Scatterers of finite dimensions and arbitrary shapes are most commonly described by spherical T-matrices, whereas cylindrical T-matrices are naturally suited for scatterers with translational symmetry along one axis, such as infinitely long cylinders or one-dimensional (1D) periodic chains of scatterers. Analytic formulas exist for spheres in the SWB and infinite cylinders in the CWB~\cite{Bohren_Wiley1983}. For other geometries, the T-matrix can be computed numerically. Reference~\citenum{asadova_jqsrt333} offers an overview of computations of spherical T-matrices using various numerical Maxwell solvers. Concerning the illumination, generally the CWB is well-suited for incident fields with fixed $k_z$, as is the case of fast electrons, and can be readily expanded to the basis of the T-matrix characterizing the scatterer using \eqsref{eq:CW_to_CW} and~(\ref{eq:CW_to_SW}).

The T-matrix method provides an elegant framework for multiple scattering in ensembles of nanostructures. Consider $N$ scatterers at positions $\vect{r}_i$ with $i=1,2,\ldots, N$, each one described by a T-matrix $\vect{T}_i$. To describe the scattered field at scatterer $i$, one has to include the scattered fields of all other scatterers as an incident field in addition to the incident field $\vect{a}$ of the fast electron, as
\begin{equation} \label{eq:p_i}
    \vect{p}_i = \vect{T}_i \left[ \vect{a}_i + \sum_{j \neq i}^N \vect{C}^{(3)} (\vect{r}_i -\vect{r}_j) \vect{p}_j \right] \, ,
\end{equation}
where $\vect{C}^{(3)} (\vect{r}_i -\vect{r}_j) = \vect{C}^{(3)}_{i,j}$ is the translation operator that converts an outgoing-wave expansion at $\vect{r}_j$ into a regular-wave expansion about $\vect{r}_i$. We refer the reader to Ref.~\cite{beutel_cpc297} for the explicit formulas of the translation coefficients comprising $\vect{C}^{(3)}_{i,j}$, as well as those of $\vect{C}^{(1)}_{i,j}$, containing the regular-to-regular-wave translation coefficients. Figure~\ref{fig:domains}(b) shows an example of a cluster with three scatterers. In a local description, the field expansions $\vect{a}_i$ and $\vect{p}_i$ are expressed in a SWB centered at the position of the $i^\mathrm{th}$ scatterer, and are valid outside the corresponding circumscribing sphere (black dashed circle), which must not overlap. By combining all field expansions in a single vector $\vect{a}_\mathrm{local}$ and $\vect{p}_\mathrm{local}$, \eqref{eq:p_i} can be recast in a form similar to \eqref{eq:p_single}, as
\begin{equation} \label{eq:p_local}
    \vect{p}_\mathrm{local} =  \vect{T}_\mathrm{local}  \, \vect{a}_\mathrm{local} \, ,
\end{equation}
with
\begin{equation} \label{eq:T_local}
\vect{T}_\mathrm{local} = [\eye - \vect{T}_\mathrm{diag} \vect{C}^{(3)}]^{-1}\vect{T}_\mathrm{diag} \, ,
\end{equation}
where $\vect{T}_\mathrm{diag}$ is a block-diagonal matrix containing the T-matrices of all particles,
\begin{align} \label{eq:Tdiag_C}
\vect{T}_\mathrm{diag} = 
    \begin{pmatrix}
        \vect{T}_1 & \zero & \cdots & \zero\\
        \zero& \vect{T}_2 & \ddots & \vdots\\
        \vdots & \ddots & \ddots & \zero\\
        \zero & \cdots & \zero & \vect{T}_N
    \end{pmatrix} \, ,
    \text{ and }
    \vect{C}^{(3)} = 
    \begin{pmatrix}
        \zero  & \vect{C}^{(3)}_{1,2} & \cdots &\vect{C}^{(3)}_{1,N}\\
        \vect{C}^{(3)}_{2,1} & \zero  & \ddots & \vdots\\
        \vdots & \ddots & \ddots & \vect{C}^{(3)}_{N-1,N}\\
        \vect{C}^{(3)}_{N, 1} & \cdots & \vect{C}^{(3)}_{N, N-1} & \zero 
    \end{pmatrix} \, .
\end{align}
Since matrix $\vect{T}_\mathrm{local}$ accounts for interactions between multiple scatterers, it can already be used to simulate CL and EELS spectra, as we demonstrate in the following section. To account for the total response from the entire cluster, one needs to convert the local T-matrix to a global matrix $\vect{T}$, where all fields are expanded around a single origin at $\vect{r}_0 = 0$. The global incident field $\vect{a}$ and scattered field  $\vect{p}$ coefficients are connected to the respective local ones via
\begin{subequations}
    \begin{align}
        \vect{a}_\mathrm{local} &= \left(\vect{C}^{(1)}_{1,0} \cdots \vect{C}^{(1)}_{N,0} \right) \vect{a} \\
        \vect{p} &= \left(\vect{C}^{(1)}_{0,1} \cdots \vect{C}^{(1)}_{0,N} \right) \vect{p}_\mathrm{local} \, , 
    \end{align}
\end{subequations}
and the global matrix is given by
\begin{align}\label{eq:T_global}
    \vect{T} = \left(\vect{C}^{(1)}_{0,1} \cdots \vect{C}^{(1)}_{0,N} \right) \vect{T}_\mathrm{local}\left(\vect{C}^{(1)}_{1,0} \cdots \vect{C}^{(1)}_{N,0} \right) \, . 
\end{align}
In Fig.~\ref{fig:domains}(b), we mark the validity domain of the global description in a SWB with the red dotted circle, which lies outside the circumscribing sphere enclosing all scatterers~\cite{beutel_cpc297}. Within this description, the impact parameter of the electron beam, i. e. the distance between the electron trajectory and the axes origin, must be larger than the radius of this circumscribing sphere. However, this constraint does not apply when operating in the local basis. Within the domain where both bases are valid, the two descriptions are equivalent.

Finally, we address structures with periodic boundaries in one dimension, such as the one depicted in Fig.~\ref{fig:domains}(c). We demonstrated in Section~\ref{subsec:theory:ebeam} that the field of the electron has a fixed wavevector component $k_z = k/\beta$ along its trajectory. Assuming a lattice $L$ composed of a unit cell with $N$ particles periodically arranged in space with pitch $\vect{\Lambda} = \Lambda\uvec{z}$ along the axis of the electron trajectory, the phase difference between unit cells is $\ex{\ii k_z \Lambda}$. Then the scattered field amplitudes assume a form similar to \eqref{eq:p_i}, but now including a sum spanning over all lattice sites, as
\begin{equation} \label{eq:p_i_periodic}
    \widetilde{\vect{p}}_i = \vect{T}_i \Bigg[ \widetilde{\vect{a}}_i + \sum_{j=1}^N \underbrace{{\sum_{\vect{\Lambda}\in L}}' \vect{C}^{(3)} (\vect{r}_i - \vect{r}_j - \vect{\Lambda}) \ex{\ii k_z \Lambda}}_{\widetilde{\vect{C}}^{(3)}_{i,j}} \,\widetilde{\vect{p}}_j \Bigg] \, ,
\end{equation}
where the primed sum denotes the omission of the self-interaction term with $\vect{\Lambda} = 0$ for $i=j$. In the local basis, and combining all scattered field expansions $\widetilde{\vect{p}}_\mathrm{local}$ analogously to \eqref{eq:p_local}, we have $\widetilde{\vect{p}}_\mathrm{local} = \vect{T}_\mathrm{local}  \widetilde{\vect{a}}_\mathrm{local} $, with $ \vect{T}_\mathrm{local} = \left[\eye - \vect{T}_\mathrm{diag} \widetilde{\vect{C}}^{(3)} \right]^{-1}\vect{T}_\mathrm{diag} $. Here, the matrix $\widetilde{\vect{C}}^{(3)}$ is analogous to \eqref{eq:Tdiag_C}, but now with the diagonal blocks generally being non-zero and containing the interaction of the particle with its periodic equivalents. Figure~\ref{fig:domains}(c) shows the validity domain of a local and a global description. The local spherical-wave expansions of the fields are valid outside the circumscribing sphere of each scatterer. The red dotted circles mark the domain of validity of a global T-matrix, which describes the scatterers of a unit cell as a single object. In a periodic arrangement, this description might be invalid due to overlapping circumscribing spheres, in which case only the local description can be employed. 

With the incident field given and the scattered field vector obtained from the T-matrix formalism, the relevant observables can now be expressed in a compact form. In the following, we derive expressions for the CL and EELS probabilities in terms of these vectors, namely in a generalized T-matrix formalism.

\subsection{Cathodoluminescence}

During the interaction of an electron beam with a scatterer, a fraction of the kinetic energy of the electron that is transferred to the optical modes of the structure is radiated back to the environment. This emission is collected in CL spectroscopy and is simulated by the respective probability $\Gamma_\mathrm{CL} (\omega) $ of collecting a photon of energy $\hbar\omega$ in the far field. The CL probability can be calculated via the power of the electromagnetic fields radiating into the far field, integrated over a surface that encloses the structure. In the SWB, the enclosing surface is that of an arbitrarily large sphere with infinitesimal solid angle $\dd \Omega = \sin{\theta}\, \dd \theta \,\dd \phi$, and the CL probability is given by~\cite{stamatopoulou_josab42} 
\begin{align}\label{eq:G_CL_s_definition}
      \Gamma_\mathrm{CL}^\mathrm{S} (\omega) = \frac{r^2}{\pi \hbar\omega} \oint \dd\Omega \, \mathrm{Re} \left\{ \vect{E}_\mathrm{sca} (\vect{r}, \omega) \times \vect{H}_\mathrm{sca}^* (\vect{r}, \omega) \right\} \cdot \uvec{r}\,. 
\end{align}
By introducing the electric field expansion given in \eqref{eq:scat_efield_general}, along with $\vect{H}_\mathrm{sca} (\vect{r}, \omega) = \vecnabla \times \vect{E}_\mathrm{sca} (\vect{r}, \omega)/(\ii\omega\mu_0)$, and in the limit $kr \to \infty$, we find 
\begin{equation} \label{eq:G_CL_s}
       \Gamma_\mathrm{CL}^\mathrm{S} (\omega) =\frac{1}{\pi \hbar\omega Z}  \frac{\vect{p}^\dagger \vect{p}}{k^2}  \,,  
\end{equation}
where $Z=\sqrt{\mu_0/\varepsilon_0}$ is the impedance in vacuum. 

In the CWB, the enclosing surface is that of a cylinder of length $L$ and an infinitesimal normal surface element $ \rho \,\dd \phi \, \dd z \,\urho$, yielding
\begin{align}\label{eq:G_CL_c_definition}
      \Gamma_\mathrm{CL}^\mathrm{C} (\omega) = \frac{\rho}{\pi \hbar\omega} \int_{0}^{2\pi} \int_0^{L}  \dd \phi\,  \dd z \, \mathrm{Re} \left\{ \vect{E}_\mathrm{sca} (\vect{r}, \omega) \times \vect{H}_\mathrm{sca}^* (\vect{r}, \omega) \right\} \cdot \urho \,. 
\end{align}
The scattered field in \eqref{eq:scat_efield_general} in the CWB may, in general, contain both propagating modes ($k_\rho \in \mathbb{R}$, where $\mathbb{R}$ is the set of real numbers) and evanescent modes ($k_\rho \in \mathbb{I}$) in the radial direction. For the latter, \eqref{eq:G_CL_c_definition} yields zero. For propagating modes and in the limit $k_\rho\rho \to \infty$, we obtain
\begin{equation} \label{eq:G_CL_c}
       \Gamma_\mathrm{CL}^\mathrm{C} (\omega) =\frac{4 L}{\pi\hbar \omega Z} \frac{\vect{p}^\dagger \vect{p}}{k}\, ,
\end{equation}
where the vector $\vect{p}$ contains the expansion coefficients of the propagating modes only. For structures that extend infinitely along the axis of the cylindrical waves, we can express the CL probability per unit length, namely \eqref{eq:G_CL_c} divided by the length $L$.

\subsection{Electron energy loss}

The measured quantity in EELS is the total electron energy loss, and is simulated by the respective probability of the electron losing energy $\hbar\omega$ while traveling against the scattered electromagnetic field, defined as
\begin{align}\label{eq:G_EELS_definition}
   \Gamma_\mathrm{EELS}(\omega) = \frac{\elc}{\pi\hbar\omega} \int_{-\infty}^{\infty} \dd t\, \mathrm{Re} \left\{ \mathrm{e}^{-\ii\omega t} \vect{v} \cdot \vect{E}_\mathrm{sca} (\vect{r}_e, \omega) \right\}\, .
\end{align}
By substituting in \eqref{eq:G_EELS_definition} the electric field expansion of \eqref{eq:scat_efield_general} in terms of spherical waves, we find 
\begin{align}\label{eq:G_EELS_s}
   \Gamma_\mathrm{EELS}^\mathrm{S}(\omega) = - \frac{ 1}{\pi\hbar\omega Z} \frac{\mathrm{Re} \{\vect{a}^\dagger \vect{p}\} }{k^2}  \, ,
\end{align}
whereas, by inserting the cylindrical wave expansion, we obtain 
\begin{align}\label{eq:G_EELS_c}
   \Gamma_\mathrm{EELS}^\mathrm{C}(\omega) = \frac{4 L}{\pi\hbar \omega Z} \frac{\mathrm{Re} \{\vect{a}^\mathrm{T} \vect{p}\} }{k} \, .
\end{align}
Similarly to the CL probability, for structures that extend infinitely along the axis of the cylindrical waves, we typically work with the EELS probability per unit length of electron trajectory, \textit{i.e.}, \eqref{eq:G_EELS_c} divided by the length $L$.

\section{Numerical implementation and examples}\label{sec:numerics}

In this section, we demonstrate the T-matrix formalism for CL and EELS simulations, as described in the previous sections, using a selected set of problems. To do so, we present here our extension of the in-house developed Python toolbox \treams, designed for electromagnetic scattering computations using the T-matrix method. The open-source code is available at \url{https://github.com/tfp-photonics/treams}, and a detailed description can be found in Ref.~\citenum{beutel_cpc297}. This new module, named \treamsebeam, uses the framework of the original code while adding the electron-beam spectroscopy functionality. More specifically, it provides the electron-beam illumination, which relies on the CWB provided by \treams, and is compatible with the operations that expand this field to the SWB. The computation of the different types of T-matrices and all relevant operations are provided by \treams. In \treamsebeam, the incident and scattered fields are instantiated as \textit{PhysicsArray} objects, which is the central class in \treams and keeps track of physically relevant information. Finally, \treamsebeam provides dedicated built-in functions for computing the total CL and EELS probabilities. The open-source code for \treamsebeam is available at \url{https://github.com/tfp-photonics/treams_ebeam}, and the documentation can be found at \url{https://tfp-photonics.github.io/treams_ebeam/welcome.html}. In the following, we provide guidelines for performing simulations with the module, designed to follow a workflow similar to that of the original code.

Within \treamsebeam, the electron beam source can be initialized with the wavenumber$k$ (denoted as \texttt{k0} in the code listings), beta factor $\beta$ (denoted as \texttt{vel}), and impact parameter $b$, as
\begin{lstlisting}[label={lst:ebeam}]
    >>> inc = treams_ebeam.ebeam(k0, vel, b) 
\end{lstlisting}
which creates the singular cylindrical wave of \eqref{eq:ebeam_efield_cwb}. To expand the electron field as a regular field in the global CWB, we can use the \texttt{expand} operator:
\begin{lstlisting}
    >>> kz = k0/vel
    >>> cwb = treams.CylindricalWaveBasis.default(kz,mmax)
    >>> inc = inc.expand(cwb, 'regular')
\end{lstlisting}
To obtain the incident field in the SWB, one must first apply the above change of basis to expand the field in terms of regular cylindrical waves, which are then converted to regular spherical waves, according to \eqref{eq:CW_to_SW}, using once again the \texttt{expand} operator:
\begin{lstlisting}
    >>> swb = treams.SphericalWaveBasis.default(lmax)
    >>> inc = inc.expand(swb)
\end{lstlisting}
Given a T-matrix, the CL and EELS probability is calculated using the expressions that correspond to each basis as
\begin{lstlisting}
    >>> cl = treams_ebeam.cl(tmatrix, inc)
    >>> eels = treams_ebeam.eels(tmatrix, inc) 
\end{lstlisting}
For these operations to be performed correctly, the incident field and the T-matrix must be expressed in the same basis, and the wavenumber must be provided in units of nm$^{-1}$. The results of the \texttt{cl} and \texttt{eels} functions depend on the basis of the T-matrix; in the CWB, they correspond to CL and EELS probability per energy in eV and per unit length in nm, while in the SWB, per energy in eV. As a final remark, we note that the built-in functions described above compute the total CL and EELS probabilities. For other observables, such as CL emission within a specified solid angle or EELS probability corresponding to a limited collection angle, one must compute the scattered field vector explicitly, and then use it to evaluate the quantity of interest. With the basic functionality established, we proceed to the examples.

\subsection{Interaction with single scatterer}

\begin{figure}[t]
    \centering
    \includegraphics[width=\linewidth]{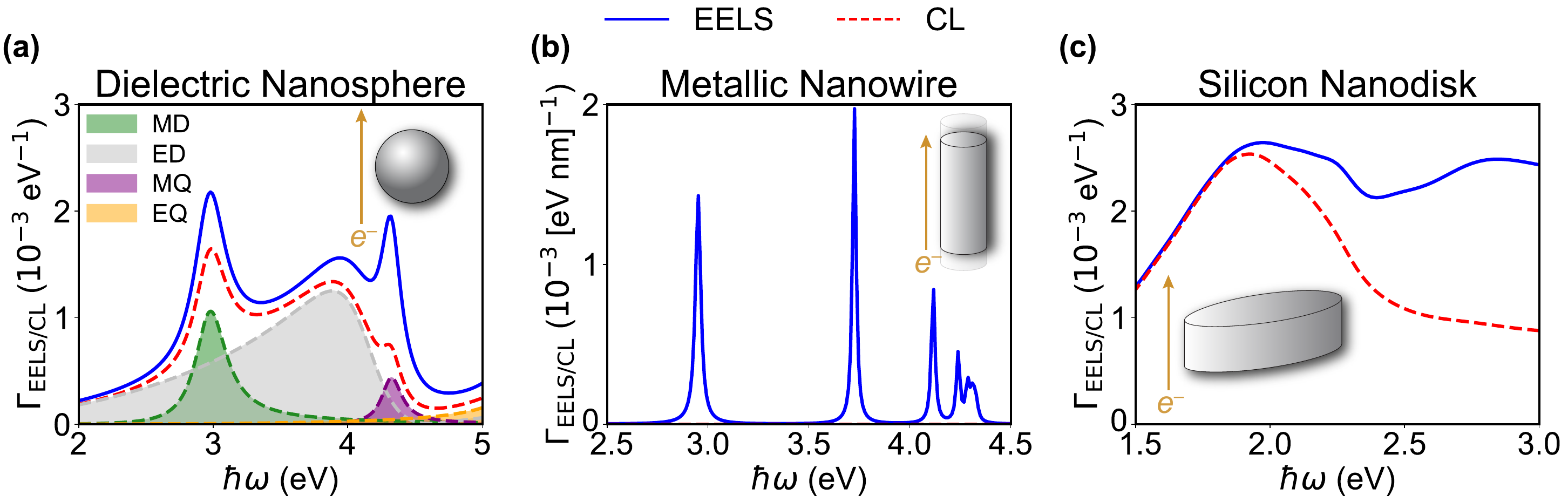}
    \caption{EELS (blue solid line) and CL probability (red dashed line) of single scatterers. (a) A dielectric nanosphere of radius $R=50$\;nm, and the decomposition of the CL probability into its magnetic dipole (MD), electric dipole (ED), magnetic quadrupole (MQ), and electric quadrupole (EQ) contributions. (b) A metallic nanowire of radius $R=50$\;nm and infinite length, and (c) an amorphous silicon elliptical nanodisk of height $h=90$\;nm, long axis $r_\mathrm{l}=286$\;nm and short axis $r_\mathrm{s}=96$\;nm. In all panels, we consider $b-R=10$\;nm and $\beta=0.7$.}
    \label{fig:single_scatterer}
\end{figure}

As our first example, we examine the CL and EELS spectra of single scatterers of different shapes and materials. Figure~\ref{fig:single_scatterer} shows the spectra for a non-dispersive dielectric sphere, an infinitely extended metallic cylinder (wire), and an elliptical nanodisk composed of amorphous silicon, considering an electron beam with $\beta=0.7$, passing at a distance $b-R = 10$\,nm from their surface.

Figure~\ref{fig:single_scatterer}(a) shows the spectra of a dielectric sphere of radius $R=50$\;nm composed of a material with nondispersive dielectric function $\varepsilon = 16+0.5\ii$. The T-matrix method is particularly well suited for classifying multipolar optical modes, since each contribution corresponds to an entry in the scattered-field expansion vector $\vect{p}$. In the energy window displayed, three peaks can be observed. The multipolar decomposition of the CL probability provided in Fig.\ref{fig:single_scatterer}(a) up to $\ell =2$ reveals that these features correspond to the magnetic dipole (MD), electric dipole (ED), and magnetic quadrupole (MQ) contributions, while modes $\ell_\mathrm{max}>4$ are negligible. The difference between the EELS and CL curve reveals that a part of the energy is absorbed, rather than radiated, due to the complex dielectric function. For this example, the T-matrix describing the sphere can be computed within \treams, using the method \texttt{TMatrix.sphere}, as
\begin{lstlisting}
    >>> tm = treams.TMatrix.sphere(lmax, k0, radius, materials)
\end{lstlisting} 
where \texttt{materials} contains the relative permittivity and permeability of the scatterer and the host medium. The T-matrices implemented in \treams are formulated within the local-response approximation. Nonlocal T-matrices can, in principle, be computed externally and imported into the program, for the study of nonlocal effects in metallic spheres, as well as bulk plasmons~\cite{christensen_acsn8}.

Besides spheres, the T-matrices of infinite cylinders can be computed with the method \texttt{TMatrixC.cylinder}:
\begin{lstlisting}
    >>> kz = k0/vel
    >>> tm = treams.TMatrixC.cylinder(kz, mmax, k0, radius, materials) 
\end{lstlisting}
Figure~\ref{fig:single_scatterer}(b) shows the spectra of a metallic wire of radius $R=50$\;nm whose dielectric function is given by a Drude model representative of silver, $\varepsilon (\omega) = \varepsilon_\infty - \omega_p^2/(\omega^2 + \ii\gamma_p\omega)$, with $\varepsilon_\infty = 3.3$, $\hbar\omega_p = 9$\;eV and $\hbar\gamma_p = 22$\;meV. The electron beam launches surface plasmon polaritons (SPPs), which are confined at the metal--air interface and propagate along it with wavenumber $k_z=k/\beta$. These take the form of purely TM cylindrical waves of different orders ($m=0,1,\ldots,m_\mathrm{max}$), and appear in the spectra as Lorentzian peaks at different energies.

For geometries other than spheres and wires, the T-matrices can be found in databases~\cite{asadova:2026} or computed using alternative numerical methods. Most of the methods mentioned in Section~\ref{sec:introduction} support this functionality. Once obtained, the T-matrix can be loaded into \treams, provided that it complies with the format specified in Ref.~\citenum{asadova_jqsrt333}. In Fig.~\ref{fig:single_scatterer}(c), we show the response of an amorphous silicon elliptical nanodisk of height $h=90$\;nm, long axis $r_\mathrm{l}=286$\;nm, and short axis $r_\mathrm{s}=96$\;nm. This structure is inspired by Ref.~\citenum{dong_lsa11}, where such disks are arranged in a 2D lattice to investigate the emergence of BICs and q-BICs. In this example, the T-matrix is computed using the BEM-based \textsc{matlab} toolbox \textsc{nanobem}~\cite{hohenester_cpc276}, with the dielectric function taken from Ref.~\citenum{dong_lsa11}, and can be found in the Daphona database~\cite{asadova:2026}. The EELS and CL spectra in Fig.~\ref{fig:single_scatterer}(c) show broad features rather than sharp Lorentzian peaks, and multipolar modes of mixed electric and magnetic type. 

\begin{figure}[t]
    \centering
    \includegraphics[width=\linewidth]{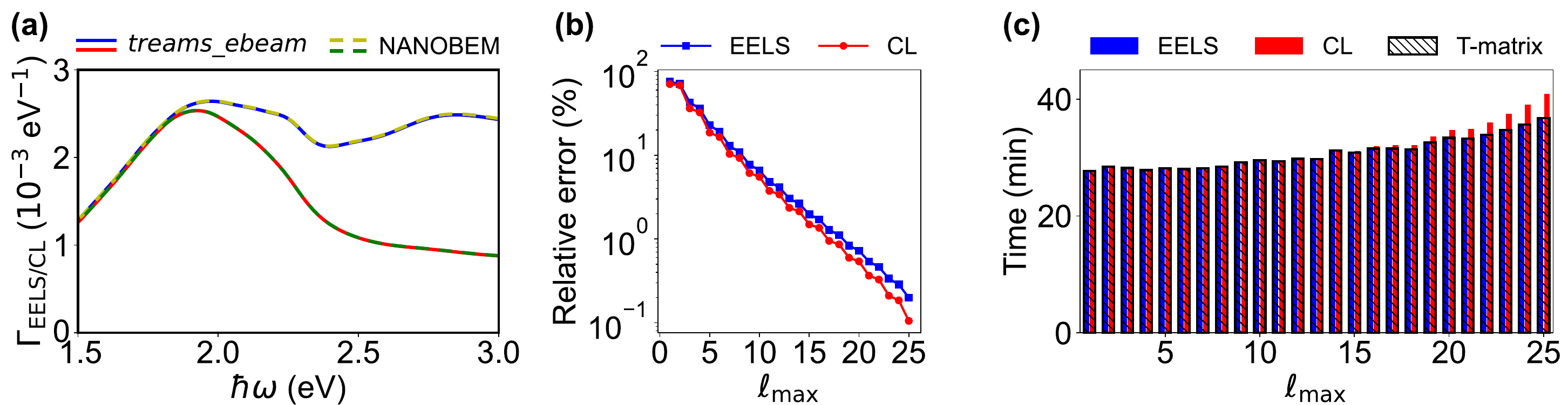}
    \caption{(a) Comparison of CL and EELS spectra of the elliptical disk analyzed in Fig.~\ref{fig:single_scatterer}(c) computed with \treamsebeam and the \textsc{nanobem} toolbox~\cite{hohenester_cpc276} when considering a T-matrix with a multipolar order up to $\ell_\mathrm{max}=25$ in the expansion. An excellent agreement is observed. (b) Relative error between the results obtained with the \textsc{nanobem} toolbox and \treamsebeam in the CL and EELS spectra, depending on the parameter that controls the convergence in the \treamsebeam simulations. Specifically, the convergence parameter is the largest multipolar order $\ell_\mathrm{max}$ used to truncate the T-matrix. (c) Computational time versus $\ell_\mathrm{max}$ to assess the computational efficiency. The hatched area denotes the computation time required to compute the T-matrix in \textsc{nanobem}. It can clearly be seen that the largest share is taken by the computational task of computing the T-matrix. Once the T-matrix is computed, the prediction of the CL and EELS spectra only takes a tiny additional amount of time.} 
    \label{fig:disk_convergence}
\end{figure}

Figure~\ref{fig:disk_convergence}(a) compares the spectra of the elliptical nanodisk analyzed in Fig.~\ref{fig:single_scatterer}(c) as computed using \treamsebeam against the same simulations performed using the \textsc{nanobem} toolbox. For a fair comparison, the same discretization ($935$ vertices) used to generate the T-matrix is employed in the \textsc{nanobem} calculations. We find that a large number of multipoles ($\ell_\mathrm{max}=25$) is required to achieve convergence throughout the entire energy window. However, using such a large multipolar order ensures an excellent agreement among the methods. Evidently, it is more the truncation of the T-matrix that affects the agreement than the actual methodology used to predict the observable spectrum. 

To quantify this convergence more explicitly, we define a relative error as 
\begin{equation}\label{eq:rel_error}
    \mathrm{Relative~error}_{\ell_\mathrm{max}} (\%) = \frac{\int_{\omega_\mathrm{min}}^{\omega_\mathrm{max}} \dd \omega \, \left|\Gamma_\textsc{nanobem}(\omega)-\Gamma_\treamsebeam^{\ell_\mathrm{max}}(\omega)\right|}{\int_{\omega_\mathrm{min}}^{\omega_\mathrm{max}} \dd \omega \, \Gamma_\textsc{nanobem}(\omega)} 100\%\, ,
\end{equation}
where we explicitly indicate that the corresponding spectrum and the relative error are calculated depending on the multipolar order used to truncate the T-matrix. We reach a relative error of $0.1\%$ in CL and $0.2\%$ in EELS probability for the largest multipolar order considered. Such a high level of accuracy is not always required in practice. For instance, already at $\ell_\mathrm{max}=14$ the relative error remains below $3\%$, which would be sufficient for many applications, including the simulation of experimental measurements. 

 Finally, as shown in Fig.~\ref{fig:disk_convergence}(c), where we compare the computational time to obtain the corresponding spectra depending on the multipolar order, we see that the computational time is dominated by the time that is spent on generating the T-matrix. Consequently, for single scatterers, \treamsebeam largely inherits both the accuracy and computational efficiency of the method used to compute the T-matrix.

The computational advantage of the T-matrix method becomes apparent when considering systems composed of multiple scatterers. Given the T-matrix of individual scatterers, \treams provides built-in methods to compute the T-matrix of periodic and aperiodic clusters, following the formulation described in Section~\ref{subsec:theory:ebeam_nanostructure}. From the T-matrix of a simple or complex unit cell, the T-matrices of periodic nanoparticle arrays in one, two, and three dimensions are readily constructed by employing the highly efficient and fast-converging Ewald method~\cite{beutel_cpc297}. We next demonstrate simulations of CL and EELS in these scenarios.

\subsection{Interaction with periodic chain of nanodisks}
\label{subsec:numerics:chain}

\begin{figure}[t]
    \centering
    \includegraphics[width=\linewidth]{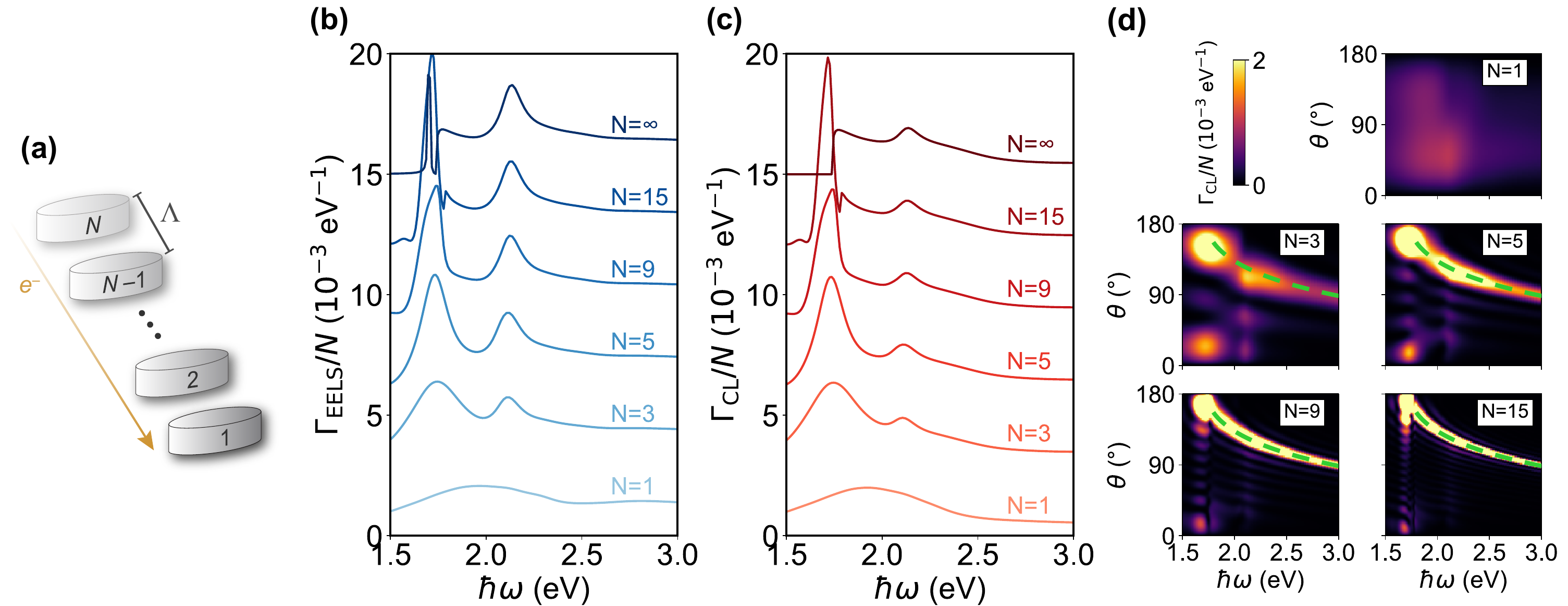}
    \caption{(a) Schematic of the system under study: a chain of amorphous silicon elliptical nanodisks excited by an electron beam with parameters $\beta=0.7$ and $b-R=7$\;nm. (b) EELS and (c) CL probabilities per nanoparticle for chains with $N =1,3,5,9,15,\infty$ nanoparticles. Consecutive curves are vertically offset by $3\cdot 10^{-3}$\;eV$^{-1}$ for readability. (d) CL probability per nanoparticle versus azimuthal angle $\theta$ for the different cases of the finite chain. The dashed green lines trace the angles at which Smith-Purcell radiation is emitted in the first diffraction order.}
    \label{fig:1d_chain}
\end{figure}

In this section, we demonstrate CL and EELS simulations of a finite and an infinite periodic chain of nanoparticles aligned in parallel to the electron trajectory. Periodic nanostructures are key building blocks for nanophotonic devices, because they support lattice interactions and resonances with enhanced quality factors, including plasmonic surface lattice resonances, BICs, and q-BICs. In the context of free electron--matter interactions, such configurations are widely studied as sources of Smith–Purcell radiation~\cite{garciadeabajo_pre61}.

As a representative example, we arrange the amorphous silicon elliptical nanodisks introduced in Fig.~\ref{fig:single_scatterer}(c) in a chain with pitch $\Lambda = 294$\;nm along the z-axis, \textit{i.e.}, parallel to the electron beam, as illustrated in Fig.~\ref{fig:1d_chain}. Figures~\ref{fig:1d_chain}(b) and \ref{fig:1d_chain}(c) show the EELS and CL probability per nanoparticle, respectively, for finite chains containing $N=1,3,5,9,$ and $15$ nanoparticles, as well as for an infinite chain ($N=\infty$). 

Starting from the T-matrix \texttt{disk} of a single object, the local T-matrix of \eqsref{eq:T_local}-(\ref{eq:Tdiag_C}) of the finite chain, namely the cluster, containing multiple-scattering interactions among all objects, is computed in \treams as:
\begin{lstlisting}
    >>> tm = treams.TMatrix.cluster([disk]*N, positions).interaction.solve()
\end{lstlisting} 
To compute the CL and EELS probabilities, the incident electron field must first be expressed in the CWB at the positions of the nanoparticles as regular waves, and subsequently expanded into the SWB of the T-matrix. The outputs of \texttt{treams\_ebeam.cl(tm, inc)} and \texttt{treams\_ebeam.eels(tm, inc)} then yield the respective total probabilities, and divided by $N$ they yield probability per unit cell, here per nanoparticle. Although it is possible to expand the T-matrix of the cluster from the local to the global basis, this operation will limit the validity of the computation to the domain outside the circumscribing sphere enclosing all scatterers, which is not compatible with the chosen impact parameter. Consequently, the EELS and CL calculations are performed using the local T-matrix \texttt{tm}. 

For the infinite chain, the local T-matrix for a given lattice pitch and longitudinal component of the wavevector $k_z = k/\beta$ (where $\hbar k_z$ is the momentum transferred from the electron beam to the chain) is obtained by solving the interaction between periodic unit cells:
\begin{lstlisting}
    >>> lattice = treams.Lattice(pitch)
    >>> chain = disk.latticeinteraction.solve(lattice, kz)
\end{lstlisting} 
The resulting T-matrix \texttt{chain} is expressed in the SWB and can be used directly to calculate EELS probability per unit cell, using the \texttt{eels} method as described so far. Alternatively, the chain of periodic spherical waves transforms to a scattered cylindrical wave, thereby converting the T-matrix from a local SWB to a CWB with a single origin:
\begin{lstlisting}
    >>> cwb = treams.CylindricalWaveBasis.diffr_orders(kz, lmax, lattice, bmax)
    >>> chain_tmc = treams.TMatrixC.from_array(chain, cwb)
\end{lstlisting} 
Here, the parameter $b_\mathrm{max} = 2\pi n/\Lambda$ determines the diffraction orders $n$ included in the simulation. In the CWB, \eqsref{eq:G_CL_c} and (\ref{eq:G_EELS_c}) yield the probabilities per unit length in nm, while the corresponding probabilities per unit cell are obtained by multiplying by the lattice pitch. For more information regarding the computation of T-matrices for clusters and periodic arrays of simple and complex unit cells, we refer the reader to the documentation webpage of \treams at  \url{https://github.com/tfp-photonics/treams}, and the respective publication in Ref.~\citenum{beutel_cpc297}.

As shown in Figs.~\ref{fig:1d_chain}(b) and \ref{fig:1d_chain}(c), a lattice resonance emerges at approximately $1.7$\;eV due to the collective interactions among the nanodisks, becoming progressively sharper with each additional nanoparticle. In the infinite-chain limit (darkest blue and red curves), the radiative emission of this resonance is fully suppressed, as evidenced by the difference between CL and EELS spectra. In this regime, the CL emission arises from Smith--Purcell radiation, and photons of each energy are emitted in a cone of angle $\theta_n = \cos^{-1} (1/\beta - b_\mathrm{max}(n)/k)$ for each diffraction order $n$; here we set $n=1$. Figure~\ref{fig:1d_chain}(d) shows the angular dependence of the CL emission for the different cases of the finite chain. While the emission from a single nanodisk is essentially omnidirectional, increasing the number of scatterers in the finite chain leads to progressively higher directionality, approaching the characteristic angular emission pattern of Smith–Purcell radiation, indicated with dashed green lines in the panels for $N= 3, 5, 9,$ and $15$. The nodes observed in the angular spectra correspond to the number of unit cells in the chain.

\begin{figure}[t]
    \centering
    \includegraphics[width=0.95\linewidth]{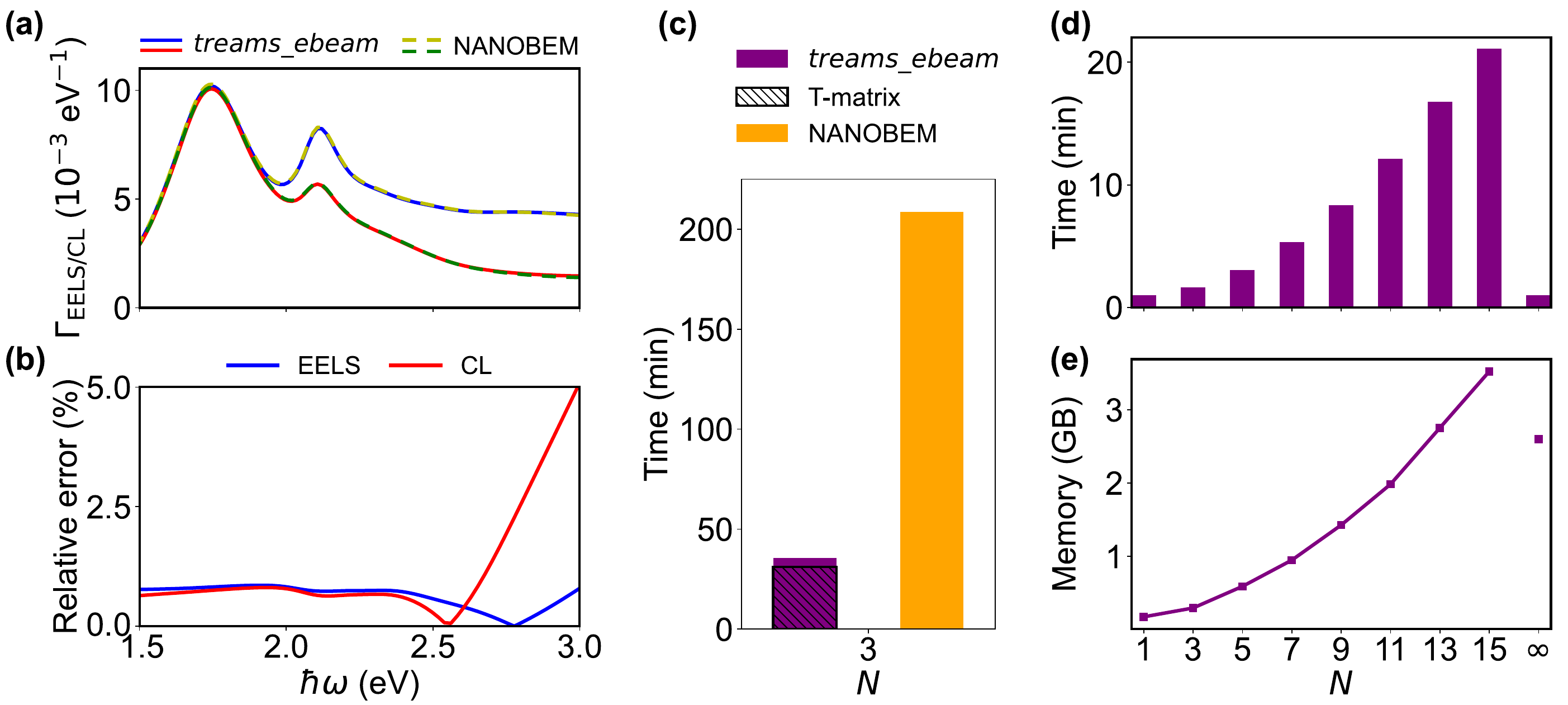}
    \caption{(a) Comparison of the EELS and CL probability of the elliptical disk cluster with $N=3$ analyzed in Fig.~\ref{fig:1d_chain} computed with \treamsebeam and the \textsc{nanobem} toolbox. (b) Respective relative error between the CL and EELS spectra computed by the two methods, given by $\left|\Gamma_\textsc{nanobem}(\omega)-\Gamma_\treamsebeam(\omega)\right|/\Gamma_\textsc{nanobem}(\omega)\cdot 100\%$. (c) Computational time for generating the spectra shown in panel (a), showcasing the computational efficiency of the T-matrix approach in systems comprising multiple scatterers. The hatched area denotes the computational time required to compute the T-matrix of the unit cell in \textsc{nanobem}. (d) Computational time and (e) peak memory consumption versus number of scatters $N$ computed at a single energy point ($\hbar\omega=1.77$~eV) in simulations performed with \treamsebeam. The results were generated in a computer cluster with characteristics: Intel Xeon E5-2630 v4 CPU (40 cores, 80 threads, 2.20 GHz base clock).}
    \label{fig:1d_chain_errors}
\end{figure}

To showcase the computational advantage of the T-matrix approach for systems comprising multiple scatterers, Figs.~\ref{fig:1d_chain_errors}(a), (b), and (c) compare \treamsebeam and \textsc{nanobem} simulations for the cluster analyzed in Fig.~\ref{fig:1d_chain} with $N=3$ nanoparticles. Figure~\ref{fig:1d_chain_errors}(a) shows good agreement between the compared spectra, with the relative error remaining below $5\%$ across the entire energy window, as shown in panel~\ref{fig:1d_chain_errors}(b). In the \treamsebeam simulations, we employed the same multipolar truncation as in Fig.~\ref{fig:1d_chain}, i.e., $\ell_\mathrm{max} =14$, which is sufficient to reproduce the \textsc{nanobem} spectra with good accuracy. Figure~\ref{fig:1d_chain_errors}(c) showcases that the advantage of the T-matrix approach is that it scales more favorably with the number of scatters. It must be noted that no special techniques were employed to accelerate the \textsc{nanobem} simulations. Importantly, in the \treamsebeam simulations, most of the computational effort is devoted to computing the T-matrix of the unit cell [hatched area in panel~\ref{fig:1d_chain_errors}(c)], whereas extracting the interaction T-matrix and evaluating the spectra require comparatively little additional time. Naturally, increasing the cutoff parameter $\ell_\mathrm{max}$ improves the agreement between the methods, up to a point where numerical inaccuracies contained in the T-matrix begin to distort the spectra. As the number of scatterers increases, further refinement of this parameter becomes necessary, which in turn leads to a moderate increase in computational time, as shown in Fig.~\ref{fig:disk_convergence}(c).

In Figs.~\ref{fig:1d_chain_errors}~(d) and (e), we find that the computational cost of the simulations performed with \treamsebeam scales quadratically with the number of scatterers $N$ in the case of a finite cluster. In contrast, the simulation of the infinite periodic array ($N=\infty$) is significantly more efficient, owing to the Ewald method employed for the evaluation of the lattice sums. The computational time and peak memory consumption reported here concern the evaluation of a single spectral point, specifically at $\hbar\omega=1.77$\;eV, close to where the lattice resonance is formed for large $N$. In all calculations, we fix the value of the multipole truncation at $\ell_\mathrm{max}=14$, in order to isolate the effect of system size on the computational scaling.

\subsection{Interaction with finite 2D periodic array of nanospheres}

\begin{figure}[t]
    \centering
    \includegraphics[width=\linewidth]{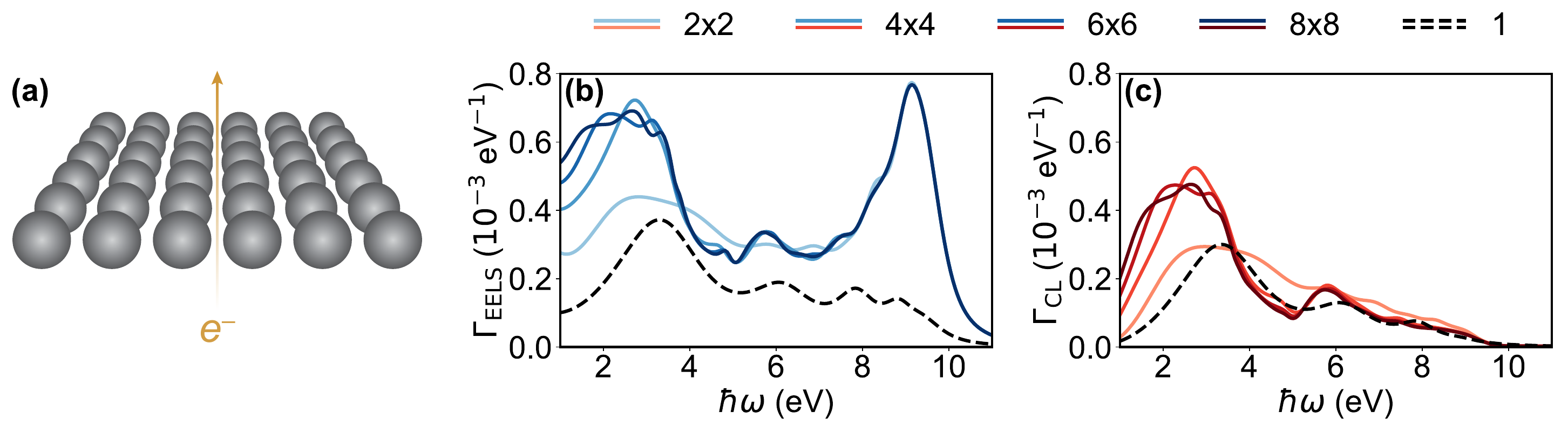}
    \caption{(a) Schematic of the system under study: a finite 2D array of aluminum nanospheres is excited by an electron beam that passes transversely through its center with $\beta=0.7$. (b) EELS and (c) CL probabilities for arrays consisting of $N = 2, 4, 6, 8$ nanoparticles per side. The black, dashed curves show the spectra of a single aluminum nanosphere positioned at the same distance from the electron trajectory as the nearest nanoparticle in the corresponding array.}
    \label{fig:2d_cluster}
\end{figure}

As a final example, we explore the system shown in Fig.~\ref{fig:2d_cluster}(a): an electron beam passing transversely through the center of a cluster of aluminum nanospheres arranged in a finite 2D array. All nanoparticles have radius $R=50$\;nm, and permittivity given by the Drude model with $\varepsilon_\infty = 1$, $\hbar\omega_p = 15$\;eV and $\hbar\gamma_p = 1.06$\;eV, while the square lattice pitch has a length of $\Lambda=120$\;nm. The T-matrix of each cluster is computed following similar steps as for the finite chains presented in Section~\ref{subsec:numerics:chain}, for maximum multipole order $\ell_\mathrm{max} = 6$.

In this case, lattice interactions can form along multiple directions. In Figs.~\ref{fig:2d_cluster}(b) and \ref{fig:2d_cluster}(c), we present the EELS (blue curves) and CL spectra (red curves), respectively, for different numbers of nanoparticles per side, compared to those of a single nanoparticle, positioned at the same distance from the electron trajectory as the nearest nanoparticle in the corresponding array (black dashed curves). In the EELS spectra of Fig.~\ref{fig:2d_cluster}(b) for the single nanoparticle, we observe four Lorentzian peaks, three corresponding to $\ell =1,2,3$, and one formed by overlapping modes with $\ell=4-6$. Once arranged in the cluster, these features split into multiple peaks, due to bonding and antibonding hybridization of plasmonic modes among the nanoparticles. In the curve corresponding to an array with two nanoparticles ($N=2$) per side, this splitting of the electric dipole mode is not resolved because the peaks lie very close in energy, appearing as a single broad resonance at $2.8$\;eV. Then, at $9.1$\;eV, we observe a high-intensity peak arising from the interaction between the overlapping higher-order modes. Naturally, this feature is not present in the CL spectra of Fig.~\ref{fig:2d_cluster}(c), as higher-order modes are very tightly confined on the surface of the individual nanoparticles, and generally do not radiate. 

Adding an extra two nanoparticles ($N=4$) per side results in a more intense resonance at lower energies, since now more nanoparticles are excited. Further increasing the number of nanoparticles does not yield substantial qualitative changes in the spectra, other than improved resolution of individual peaks at low energies. This is because the field produced by the electron is evanescent in the plane of the array and therefore couples weakly to nanoparticles farther from its trajectory. This coupling is even weaker for higher-order modes. Eventually, only a limited number of nanoparticles is excited and thus participate in the formation of the lattice resonance. At the same time, in the EELS spectra of Fig.~\ref{fig:2d_cluster}(b), the contribution of higher-order modes remains essentially unchanged for all sizes of arrays.

\section{Conclusions}\label{sec:conclusions}

We presented a T-matrix-based scattering formulation to model the interaction of fast electrons with single or multiple scatterers. Starting from the expansion of the field generated by a fast electron in the CWB, we related this source to the response of a nanostructure via the T-matrix formalism, and provided generalized formulas addressing finite and infinite 1D periodic arrays. Using these formulas, we derived expressions for computing the CL and EELS probabilities. We implemented the electron-beam spectroscopy functionality in the existing \treams open-source software suite for electromagnetic scattering computations as an add-on module called \treamsebeam. To demonstrate both the method and the software, we explored three representative examples: single scatterers, finite arrays, and infinite periodic arrays of scatterers, providing guidelines for using the module and performance comparisons against an alternative method. We envision this tool serving as a user-friendly resource for CL and EELS simulations. Future extensions to the framework, including GPU acceleration and AI-assisted simulations could further expand the range of physical systems that can be explored by improving computational performance and ease of use.

\section{Acknowledgements}
We thank N. Asadova, J.~D. Fischbach, K. Frizyuk, M. Gabbert, P. Garg, E. Herzog, L. Rebholz, and N. Ustimenko for providing technical assistance and valuable feedback on the results.
P.~E.~S. is supported by a research grant (VIL71383) from VILLUM FONDEN.

\hfill

\bibliography{references}

@article{Fiedler:2022,
author = {Fiedler, Saskia and Stamatopoulou, P. Elli and Assadillayev, Artyom and Wolff, Christian and Sugimoto, Hiroshi and Fujii, Minoru and Mortensen, N. Asger and Raza, S. and Tserkezis, Christos},
title = {Disentangling Cathodoluminescence Spectra in Nanophotonics: Particle Eigenmodes vs Transition Radiation},
journal = {Nano Lett.},
volume = {22},
number = {6},
pages = {2320-2327},
year = {2022},
doi = {10.1021/acs.nanolett.1c04754}
}

@article{Pramassing:2021,
  title = {Electron energy loss spectroscopy on freestanding perforated gold films},
  author = {Pr\"amassing, M. and Kiel, T. and Irsen, S. and Busch, K. and Linden, S.},
  journal = {Phys. Rev. B},
  volume = {103},
  issue = {11},
  pages = {115403},
  year = {2021},
  doi = {10.1103/PhysRevB.103.115403}
}

@article{Husnik:2013,
title = {Comparison of electron energy-loss and quantitative optical spectroscopy on individual optical gold antennas},
author = {Husnik, M. and von Cube, F. and Irsen, S. and Linden, S. and  Niegemann, J. and Busch, K. and  Wegener, M.},
pages = {241-245},
volume = {2},
number = {4},
journal = {Nanophotonics},
doi = {doi:10.1515/nanoph-2013-0031},
year = {2013}
}

@article{Polman:2019,
  title = {Electron-beam spectroscopy for nanophotonics},
  author = {Polman, A. and Kociak, M. and Garc{\'i}a de Abajo, F. J.},
  journal = {Nat. Mater.},
  volume = {18},
  issue = {11},
  pages = {1158-1171},
  year = {2019},
  doi = {10.1038/s41563-019-0409-1}
}

@article{kociak_um176,
  title={Cathodoluminescence in the scanning transmission electron microscope},
  author={Kociak, M and Zagonel, L F},
  journal={Ultramicroscopy},
  volume={176},
  pages={112--131},
  year={2017},
  doi={10.1016/j.ultramic.2017.03.014}
}

@article{coenen_apr4,
    author = {Coenen, T. and Haegel, N. M.},
    title = "{Cathodoluminescence for the 21st century: Learning more from light}",
    journal = {App. Phys. Rev.},
    volume = {4},
    number = {3},
    pages = {031103},
    year = {2017},
    doi = {10.1063/1.4985767}
}

@article{GarciadeAbajo:2010rmp,
  title = {Optical excitations in electron microscopy},
  author = {Garc{\'i}a de Abajo, F. J.},
  journal = {Rev. Mod. Phys.},
  volume = {82},
  issue = {1},
  pages = {209-275},
  year = {2010},
  doi = {10.1103/RevModPhys.82.209}
}

@article{GarciadeAbajo:1999prb,
  title = {Relativistic energy loss and induced photon emission in the interaction of a dielectric sphere with an external electron beam},
  author = {Garc\'{i}a de Abajo, F. J.},
  journal = {Phys. Rev. B},
  volume = {59},
  issue = {4},
  pages = {3095--3107},
  year = {1999},
  doi = {10.1103/PhysRevB.59.3095}
}

@article{Yamamoto:2001,
  title = {Photon emission from silver particles induced by a high-energy electron beam},
  author = {Yamamoto, N. and Araya, K. and Garc\'{i}a de Abajo, F. J.},
  journal = {Phys. Rev. B},
  volume = {64},
  issue = {20},
  pages = {205419},
  year = {2001},
  doi = {10.1103/PhysRevB.64.205419}
}

@article{watanabe:1956,
  title={Experimental Evidence for the Collective Nature of the Characteristic Energy Loss of Electrons in Solids --{S}tudies on the Dispersion Relation of Plasma Frequency--},
  author={Hiroshi Watanabe},
  journal={J. Phys. Soc. Japan},
  volume={11},
  number={2},
  pages={112-119},
  year={1956},
  doi={10.1143/JPSJ.11.112}
}

@article{Vesseur:2007,
author = {Vesseur, Ernst Jan R. and de Waele, R and Kuttge, Martin and Polman, Albert},
title = {Direct Observation of Plasmonic Modes in {Au} Nanowires Using High-Resolution Cathodoluminescence Spectroscopy},
journal = {Nano Lett.},
volume = {7},
number = {9},
pages = {2843-2846},
year = {2007},
doi = {10.1021/nl071480w}    
}

@article{ritchie:1957,
  title = {Plasma Losses by Fast Electrons in Thin Films},
  author = {Ritchie, R. H.},
  journal = {Phys. Rev.},
  volume = {106},
  issue = {5},
  pages = {874--881},
  year = {1957},
  doi = {10.1103/PhysRev.106.874}
}

@article{lagos:2017,
  title={Mapping vibrational surface and bulk modes in a single nanocube},
  author={Lagos, Maureen J and Tr{\"u}gler, Andreas and Hohenester, Ulrich and Batson, Philip E},
  journal={Nature},
  volume={543},
  number={7646},
  pages={529--532},
  year={2017},
  doi = {10.1038/nature21699}
}

@article{hage:2019,
  title = {Phonon Spectroscopy at Atomic Resolution},
  author = {Hage, F. S. and Kepaptsoglou, D. M. and Ramasse, Q. M. and Allen, L. J.},
  journal = {Phys. Rev. Lett.},
  volume = {122},
  issue = {1},
  pages = {016103},
  year = {2019},
  doi = {10.1103/PhysRevLett.122.016103}
}

@article{coenen:2013,
  title={Resonant modes of single silicon nanocavities excited by electron irradiation},
  author={Coenen, Toon and van de Groep, Jorik and Polman, Albert},
  journal={ACS Nano},
  volume={7},
  number={2},
  pages={1689--1698},
  year={2013},
  doi={10.1021/nn3056862}
}

@article{maciel_prb102,
  title = {Probing and steering bulk and surface phonon polaritons in uniaxial materials using fast electrons: Hexagonal boron nitride},
  author = {Maciel-Escudero, C. and Kone\v{c}n\'{a}, Andrea and Hillenbrand, Rainer and Aizpurua, Javier},
  journal = {Phys. Rev. B},
  volume = {102},
  issue = {11},
  pages = {115431},
  year = {2020},
  doi = {10.1103/PhysRevB.102.115431}
}

@article{matsukata:2019,
  title={Selection and visualization of degenerate magnetic and electric multipoles up to radial higher orders by cathodoluminescence},
  author={Matsukata, Taeko and Matthaiakakis, Nikolaos and Yano, Taka-aki and Hada, Masaki and Tanaka, Takuo and Yamamoto, Naoki and Sannomiya, Takumi},
  journal={ACS Photonics},
  volume={6},
  number={9},
  pages={2320--2326},
  year={2019},
  doi = {10.1021/acsphotonics.9b00833}
}

@book{egerton:2011,
  title={Electron Energy-Loss Spectroscopy in the Electron Microscope},
  author={Egerton, Ray F},
  year={2011},
  publisher={Springer Science+Business Media},
address = {New York},
edition = {3rd edition},
  doi={10.1007/978-1-4419-9583-4}
}

@article{GarciadeAbajo:2002prb,
  title = {Retarded field calculation of electron energy loss in inhomogeneous dielectrics},
  author = {Garc\'{i}a de Abajo, F. J. and Howie, A.},
  journal = {Phys. Rev. B},
  volume = {65},
  issue = {11},
  pages = {115418},
  numpages = {17},
  year = {2002},
  doi = {10.1103/PhysRevB.65.115418}
}

@article{hohenester:2014,
  title={Simulating electron energy loss spectroscopy with the {MNPBEM} toolbox},
  author={Hohenester, Ulrich},
  journal={Comput. Phys. Commun.},
  volume={185},
  number={3},
  pages={1177--1187},
  year={2014},
  doi={10.1016/j.cpc.2013.12.010}
}

@article{matyssek:2011,
  title={Computing electron energy loss spectra with the Discontinuous {G}alerkin Time-Domain method},
  author={Matyssek, Christian and Niegemann, Jens and Hergert, Wolfram and Busch, Kurt},
  journal={Photon. Nanostruct. Fundam. Appl.},
  volume={9},
  number={4},
  pages={367--373},
  year={2011},
  doi={10.1016/j.photonics.2011.04.003}
}

@article{cao:2015,
  title={Electron energy-loss spectroscopy calculation in finite-difference time-domain package},
  author={Cao, Yang and Manjavacas, Alejandro and Large, Nicolas and Nordlander, Peter},
  journal={ACS Photonics},
  volume={2},
  number={3},
  pages={369--375},
  year={2015},
  doi={10.1021/ph500408e}
}

@article{das:2012,
  title={Probing higher order surface plasmon modes on individual truncated tetrahedral gold nanoparticle using cathodoluminescence imaging and spectroscopy combined with {FDTD} simulations},
  author={Das, Pabitra and Chini, Tapas Kumar and Pond, James},
  journal={J. Phys. Chem. C},
  volume={116},
  number={29},
  pages={15610--15619},
  year={2012},
  doi={10.1021/jp3047533}
}

@article{zabala:1989,
  title={Energy loss of electrons travelling through cylindrical holes},
  author={Zabala, Nerea and Rivacoba, Alberto and Echenique, Pedro M},
  journal={Surf. Sci.},
  volume={209},
  number={3},
  pages={465--480},
  year={1989},
  doi={10.1016/0039-6028(89)90089-7}
}

@book{Bohren_Wiley1983,
author = {Bohren, C.~F. and Huffman, D.~R.},
title = {Absorption and Scattering of Light by Small Particles},
year = {1983},
publisher = {John Wiley \& Sons},
address = {New York},
doi = {10.1002/9783527618156}
}

@article{stamatopoulou_josab42,
author = {Stamatopoulou, P.~Elli and Maciel-Escudero, Carlos and Mortensen, N.~Asger and Rockstuhl, Carsten and Tserkezis, Christos},
journal = {J. Opt. Soc. Am. B},
number = {7},
pages = {1620--1637},
title = {Analytic methods in electron energy-loss and cathodoluminescence spectroscopy of planar and spherical nanostructures: tutorial},
volume = {42},
year = {2025},
doi = {10.1364/JOSAB.559215}
}

@article{beutel_cpc297,
title = {treams -- {A} {T}-matrix-based scattering code for nanophotonics},
journal = {Comput. Phys. Commun.},
volume = {297},
pages = {109076},
year = {2024},
doi = {https://doi.org/10.1016/j.cpc.2023.109076},
author = {Beutel, Dominik and Fernandez-Corbaton, Ivan and Rockstuhl, Carsten}
}

@article{asadova_jqsrt333,
    author = {Asadova, Nigar and Achouri, Karim and Arjas, Kristian and Augui{\'e}, Baptiste and Aydin, Roland and Baron, Alexandre and Beutel, Dominik and Bodermann, Bernd and Boussaoud, Kaoutar and Burger, Sven and    others},
	doi = {10.1016/j.jqsrt.2024.109310},
	issn = {0022-4073},
	journal = {J. Quant. Spectrosc. Radiat. Transf.},
	pages = {109310},
	title = {T-matrix representation of optical scattering response: Suggestion for a data format},
	volume = {333},
	year = {2025}
}

@article{stamatopoulou_prr6,
  title = {Electron beams traversing spherical nanoparticles: Analytic and numerical treatment},
  author = {Stamatopoulou, P. Elli and Zhao, Wenhua and Rodr\'{\i}guez Echarri, \'Alvaro and Mortensen, N. Asger and Busch, Kurt and Tserkezis, Christos and Wolff, Christian},
  journal = {Phys. Rev. Res.},
  volume = {6},
  pages = {013239},
  year = {2024},
  doi = {10.1103/PhysRevResearch.6.013239}
}

@article{kepaptsoglou_nat644,
  title={Magnon spectroscopy in the electron microscope},
  author={Kepaptsoglou, Demie and Castellanos-Reyes, Jos{\'e} {\'A}ngel and Kerrigan, Adam and Alves do Nascimento, J{\'u}lio and Zeiger, Paul M and El Hajraoui, Khalil and Idrobo, Juan Carlos and Mendis, Budhika G and Bergman, Anders and Lazarov, Vlado K and others},
  journal={Nature},
  volume={644},
  number={8075},
  pages={83--88},
  year={2025},
  doi={N10.1038/s41586-025-09318-y}
}

@article{dong_lsa11,
  title={Nanoscale mapping of optically inaccessible bound-states-in-the-continuum},
  author={Dong, Zhaogang and Mahfoud, Zackaria and Paniagua-Dom{\'\i}nguez, Ram{\'o}n and Wang, Hongtao and Fern{\'a}ndez-Dom{\'\i}nguez, Antonio I and Gorelik, Sergey and Ha, Son Tung and Tjiptoharsono, Febiana and Kuznetsov, Arseniy I and Bosman, Michel and  Yang, Joel K. W.},
  journal={Light Sci. Appl.},
  volume={11},
  number={1},
  pages={20},
  year={2022},
  doi={10.1038/s41377-021-00707-2}
}

@article{smith_purcell,
  title = {Visible Light from Localized Surface Charges Moving across a Grating},
  author = {Smith, S. J. and Purcell, E. M.},
  journal = {Phys. Rev.},
  volume = {92},
  issue = {4},
  pages = {1069--1069},
  year = {1953},
  doi = {10.1103/PhysRev.92.1069}
}

@article{pendry_martinmoreno,
  title={Energy loss by charged particles in complex media},
  author={Pendry, John B and Martin-Moreno, L},
  journal={Phys. Rev. B},
  volume={50},
  number={8},
  pages={5062},
 year = {1994},
 doi = {10.1103/PhysRevB.50.5062}
}

@article{garciadeabajo_prl91,
  title = {Cherenkov Effect as a Probe of Photonic Nanostructures},
  author = {Garc\'{\i}a de Abajo, F. J. and Pattantyus-Abraham, A. G. and Zabala, N. and Rivacoba, A. and Wolf, M. O. and Echenique, P. M.},
  journal = {Phys. Rev. Lett.},
  volume = {91},
  issue = {14},
  pages = {143902},
  numpages = {4},
  year = {2003},
  month = {Sep},
  publisher = {American Physical Society},
  doi = {10.1103/PhysRevLett.91.143902}
}

@article{chen_ol48,
author = {Chen, Fei and Fan, Ren-Hao and Chen, Jun-Xuan and Liu, Yu and Hou, Ben-Qi and Peng, Ru-Wen and Wang, Mu},
journal = {Opt. Lett.},
number = {8},
pages = {2002--2005},
title = {Tuning Smith--Purcell radiation by rotating a metallic nanodisk array},
volume = {48},
year = {2023},
doi = {10.1364/OL.484324}
}

@article{kichigin_jpcc127,
author = {Kichigin, Alexander A. and Yurkin, Maxim A.},
title = {Simulating Electron Energy-Loss Spectroscopy and Cathodoluminescence for Particles in Arbitrary Host Medium Using the Discrete Dipole Approximation},
journal = {J. Phys. Chem. C},
volume = {127},
number = {8},
pages = {4154-4167},
year = {2023},
doi = {10.1021/acs.jpcc.2c06813}
}

@article{maciel_nc14,
  title={Probing optical anapoles with fast electron beams},
  author={Maciel-Escudero, Carlos and Yankovich, Andrew B and Munkhbat, Battulga and Baranov, Denis G and Hillenbrand, Rainer and Olsson, Eva and Aizpurua, Javier and Shegai, Timur O},
  journal={Nat. Commun.},
  volume={14},
  number={1},
  pages={8478},
  year={2023},
  doi={10.1038/s41467-023-43813-y}
}

@article{garciadeabajo_pre61,
  title={Smith-Purcell radiation emission in aligned nanoparticles},
  author={Garc{\'\i}a de Abajo, F.~J.},
  journal={Phys. Rev. E},
  volume={61},
  number={5},
  pages={5743},
  year={2000},
  doi = {10.1103/PhysRevE.61.5743}
}

@article{rodriguez_prb111,
  title={Relativistic electron energy-loss spectroscopy in cylindrical waveguides and holes},
  author={Rodr{\'\i}guez Echarri, {\'A}lvaro and Zhao, Wenhua and Busch, Kurt and Garc{\'\i}a de Abajo, F Javier},
  journal={Phys. Rev. B},
  volume={111},
  number={20},
  pages={205436},
  year={2025},
   doi = {10.1103/PhysRevB.111.205436}
}

@article{walsh_pmb63,
  title={An analytical expression for the energy loss of fast electrons travelling parallel to the axis of a cylindrical interface},
  author={Walsh, C.~A.},
  journal={Philos. Mag. B},
  volume={63},
  number={5},
  pages={1063--1078},
  year={1991},
  doi={10.1080/13642819108207585}
}

@article{huang_jmmct4,
  author={Huang, Huanting and Tsang, Leung and Colliander, Andreas and Yueh, Simon H.},
  journal={IEEE J. Multiscale Multiphysics Comput. Tech.}, 
  title={Propagation of Waves in Randomly Distributed Cylinders Using Three-Dimensional Vector Cylindrical Wave Expansions in Foldy–Lax Equations}, 
  year={2019},
  volume={4},
  pages={214-226},
  doi={10.1109/JMMCT.2019.2948022}}

@article{han_joapao10,
doi = {10.1088/1464-4258/10/01/015006},
year = {2007},
volume = {10},
number = {1},
pages = {015006},
author = {Han, Guoxia and Han, Yiping and Zhang, Huayong},
title = {Relations between cylindrical and spherical vector wavefunctions},
journal = {J. Opt. A: Pure Appl. Opt.}
}

@article{fernandez_corbaton_apr4,
author = {Fernandez-Corbaton, Ivan},
title = {An Algebraic Approach to Light–Matter Interactions},
journal = {Adv. Phys. Res.},
volume = {4},
number = {1},
pages = {2400088},
doi = {https://doi.org/10.1002/apxr.202400088},
year = {2025}
}

@article{rasoul_aom7,
author = {Alaee, Rasoul and Rockstuhl, Carsten and Fernandez-Corbaton, Ivan},
title = {Exact Multipolar Decompositions with Applications in Nanophotonics},
journal = {Adv. Opt. Mater.},
volume = {7},
number = {1},
pages = {1800783},
doi = {https://doi.org/10.1002/adom.201800783},
year = {2019}
}

@article{waterman_piee53,
  author={Waterman, P.~C.},
  journal={Proc. IEEE}, 
  title={Matrix formulation of electromagnetic scattering}, 
  year={1965},
  volume={53},
  number={8},
  pages={805-812},
  doi={10.1109/PROC.1965.4058}}

@article{mishchenko_jqsrt111,
  author={Mishchenko, Michael I and Travis, Larry D and Mackowski, Daniel W},
	doi = {10.1016/j.jqsrt.2010.01.030},
	journal = {J. Quant. Spectrosc. Radiat. Transf.},
	number = {11},
	pages = {1700–1703},
	title = {T-matrix method and its applications to electromagnetic scattering by particles: {A} current perspective},
	volume = {111},
	year = {2010}
}

@article{mishchenko_jqsrt88,
author = {Mishchenko, Michael I and Videen, Gorden and Babenko, Victor A and Khlebtsov, Nikolai G and Wriedt, Thomas},
	doi = {10.1016/j.jqsrt.2004.05.002},
	journal = {J. Quant. Spectrosc. Radiat. Transf.},
	number = {1},
	pages = {357–406},
	title = {T-matrix theory of electromagnetic scattering by partciles and its applications: a comprehensive reference database},
	volume = {88},
	year = {2004}
}

@article{mackowski_josaa13,
  title={Calculation of the {T} matrix and the scattering matrix for ensembles of spheres},
  author={Mackowski, Daniel W and Mishchenko, Michael I},
  journal={J. Opt. Soc. Am. A},
  volume={13},
  number={11},
  pages={2266--2278},
  year={1996},
  doi={10.1364/JOSAA.13.002266}
}

@article{xu_ao34,
  title={Electromagnetic scattering by an aggregate of spheres},
  author={Xu, Yu-lin},
  journal={Appl. Opt.},
  volume={34},
  number={21},
  pages={4573--4588},
  year={1995},
  doi = {10.1364/AO.34.004573}
}

@article{govyadinov_nc8,
  title={Probing low-energy hyperbolic polaritons in van der Waals crystals with an electron microscope},
  author={Govyadinov, Alexander A and Kone{\v{c}}n{\'a}, Andrea and Chuvilin, Andrey and V{\'e}lez, Sa{\"u}l and Dolado, Irene and Nikitin, Alexey Y and Lopatin, Sergei and Casanova, F{\`e}lix and Hueso, Luis E and Aizpurua, Javier and others},
  journal={Nat. Commun.},
  volume={8},
  number={1},
  pages={95},
  year={2017},
  doi={10.1038/s41467-017-00056-y}
}

@article{GarciadeAbajo:2021acsp,
    author = {Garc{\'i}a de Abajo, F. J. and Di Giulio, Valerio},
    title = {Optical Excitations with Electron Beams: Challenges and Opportunities},
    journal = {ACS Photonics},
    volume = {8},
    number = {4},
    pages = {945-974},
    year = {2021},
    doi = {10.1021/acsphotonics.0c01950}
}

@article{asadova:2026,
title={T-matrix database to promote information-driven research in nanophotonics},
  author={Asadova, Nigar and Boussaoud, Kaoutar and Meyer, J{\"o}rg and Tristram, Frank and Rockstuhl, Carsten},
  journal={Optical Materials Express},
  volume={16},
  number={6},
  pages={1534--1550},
  year={2026},
  publisher={Optica Publishing Group},
  doi = {you10.1364/OME.592265}
}

@book{arfken_weber,
  author = {Arfken, G. B. and Weber, H. J. and Harris, F. E.},
  publisher = {Academic Press},
  title = {Mathematical Methods for Physicists: A Comprehensive Guide},
  year = {2012},
  edition = {7th edition},
  doi = {10.1016/C2009-0-30629-7}
}

@book{mishchenko_2002,
  title={Scattering, absorption, and emission of light by small particles},
  author={Mishchenko, Michael I and Travis, Larry D and Lacis, Andrew A},
  year={2002},
  publisher={Cambridge university press},
  isbn={	9780521782524}
}

@article{bertrand_josaa37,
author = {Bertrand, Maxime and Devilez, Alexis and Hugonin, Jean-Paul  and Lalanne, Philippe and Vynck, Kevin},
journal = {J. Opt. Soc. Am. A},
pages = {70--83},
title = {Global polarizability matrix method for efficient modeling of light scattering by dense ensembles of non-spherical particles in stratified media},
volume = {37},
year = {2020},
doi = {10.1364/JOSAA.37.000070}
}

@article{lamprianidis_jqsrt296,
title = {Transcending the Rayleigh Hypothesis with multipolar sources distributed across the topological skeleton of a scatterer},
journal = {J. Quant. Spectrosc. Radiat. Transf.},
volume = {296},
pages = {108455},
year = {2023},
doi = {https://doi.org/10.1016/j.jqsrt.2022.108455},
author = {Lamprianidis, Aristeidis~G. and Rockstuhl, Carsten and Fernandez-Corbaton, Ivan},
}

@article{egel_oe24,
author = {Egel, Amos and Theobald, Dominik and Donie, Yidenekachew and Lemmer, Uli and Gomard, Guillaume},
journal = {Opt. Express},
number = {22},
pages = {25154--25168},
title = {Light scattering by oblate particles near planar interfaces: on the validity of the {T}-matrix approach},
volume = {24},
year = {2016},
doi = {10.1364/OE.24.025154}
}

@article{christensen_acsn8,
  title={Nonlocal response of metallic nanospheres probed by light, electrons, and atoms},
  author={Christensen, Thomas and Yan, Wei and Raza, S{\o}ren and Jauho, Antti-Pekka and Mortensen, N Asger and Wubs, Martijn},
  journal={ACS nano},
  volume={8},
  number={2},
  pages={1745--1758},
  year={2014},
  doi={10.1021/nn406153k}
}

@article{hohenester_cpc276,
title = {Nanophotonic resonance modes with the nanobem toolbox},
journal = {Comp. Phys. Commun.},
volume = {276},
pages = {108337},
year = {2022},
doi = {https://doi.org/10.1016/j.cpc.2022.108337},
author = {Hohenester, Ulrich and Reichelt, Nikita and Unger, Gerhard}
}

\end{document}